\documentclass[linenumbers]{aastex631}
\usepackage[utf8]{inputenc}
\usepackage{graphicx}
\usepackage{amsmath,amsfonts}

\usepackage{url}

\definecolor{Red}{RGB}{212,28,48}
\definecolor{Green}{RGB}{81,153,74}
\definecolor{Blue}{RGB}{47,151,255}
\definecolor{Purple}{RGB}{200,100,255}

\shorttitle{Intensity-sensitive quality assessment of extended sources in astronomical images}
\shortauthors{Li et al.}
\graphicspath{{./}{figures/}}

\begin{document}

\title{Intensity-sensitive quality assessment of extended sources in astronomical images}

\author{Xiaotong Li}
\affiliation{Oxford e-Research Centre, Department of Engineering Science, University of Oxford, Oxford, UK}

\author{Karel Ad\'{a}mek}
\affiliation{Oxford e-Research Centre, Department of Engineering Science, University of Oxford, Oxford, UK}

\author{Wesley Armour\footnote{wes.armour@oerc.ox.ac.uk}}
\affiliation{Oxford e-Research Centre, Department of Engineering Science, University of Oxford, Oxford, UK}
 
\begin{abstract}

Radio astronomy studies the Universe by observing the radio emissions of celestial bodies. Different methods can be used to recover the sky brightness distribution (SBD), which describes the distribution of celestial sources from recorded data, with the output dependent on the method used. Image quality assessment (IQA) indexes can be used to compare the differences between restored SBDs produced by different image reconstruction techniques to evaluate the effectiveness of different techniques. However, reconstructed images (for the same SBD) can appear to be very similar, especially when observed by the human visual system (HVS). Hence current structural similarity methods, inspired by the HVS, are not effective. In the past, we have proposed two methods to assess point source images, where low amounts of concentrated information are present in larger regions of noise-like data. But for images that include extended source(s), the increase in complexity of the structure makes the IQA methods for point sources over-sensitive since the important objects cannot be described by isolated point sources. Therefore, in this article we propose augmented Low-Information Similarity Index (augLISI), an improved version of LISI, to assess images including extended source(s). Experiments have been carried out to illustrate how this new IQA method can help with the development and study of astronomical imaging techniques. Note that although we focus on radio astronomical images herein, these IQA methods are also applicable to other astronomical images, and imaging techniques.

\end{abstract}


\section{Introduction}
\label{section1}
Image quality assessment (IQA) of natural images allows not only for qualitative comparisons of images, but IQA methods are also used for optimising image processing algorithms \citep{Ding2021} and for assessing the quality of image compression, image restoration, and image denoising algorithms, among many other uses \citep{BAKUROV2022116087}. 

Among numerous IQA metrics, Structural SIMilarity (SSIM; \citealt{SSIM1,SSIM12,SSIM13}) and its variants are frequently used \citep{BAKUROV2022116087}. SSIM aims to incorporate characteristics of the Human Visual System (HVS) and assess the similarity between two input images by combining luminance, contrast, and structure comparisons. 

However, radio astronomical (RA) images tend to be low-information, where the important information (celestial sources) is contained in a small proportion of the image whilst the rest of the image is noise-like. When assessing these low-information images using the HVS, which is most sensitive to phase information \citep{HVS}, the differences between images tends not to be very visible. Consequently, metrics like SSIM, which seek to incorporate HVS characteristics, perform poorly on sparse astronomical images \citep{Wider}. 

Some current variants of SSIM attach different importance to different information within an image. However, they are all insufficient to focus on the difference in the high-intensity (informative) part of RA images. Specifically, the weighting process in gradient based SSIMs (G-SSIMs; \citealt{SSIM5}) is based on edge and smoothness; fixation- (F-), percentile- (P-), and PF-SSIMs \citep{SSIM8} aim at weighting by visual importance, while the differences in low-information images may be impalpable by glancing; information content weighted SSIM (IW-SSIM; \citealt{SSIM10}) focuses on perceptual information content; and complex wavelet SSIM (CW-SSIM; \citealt{SSIM3,SSIM2}) aims to deal with structural distortions in the complex wavelet domain, while the information that we are interested in may not relate to a structural issue.

We have developed two intensity-sensitive IQA indexes in \citealt{Wider} to assess low-information images, they are InTensity Weighted SSIM index (ITW-SSIM) and Low-Information Similarity Index (LISI). In this work, we introduce an augmented version of LISI (augLISI), which further develops one of the two intensity-sensitive IQA indexes that we have developed to assess low-information images \citep{Wider}. Our new low-information IQA method, augLISI, improves the performance of LISI on radio astronomical images with extended sources. Our method allows us to:
\begin{itemize}
    \item{Quantitatively compare different implementation approaches for the \textbf{same} algorithm, see Section \ref{41};}
    \item{Quantitatively define the difference between \textbf{different} algorithms, see Section \ref{42};}
    \item{Jointly characterising the differences in different part of the source(s) structure without setting an empirical threshold or extracting sources, see Section \ref{44}.}
\end{itemize}

There are different comparison techniques that are suited for radio astronomical images, but all focus on source extraction, such as Quality Assessment Tool \citep{currentQA} which is used for image quality assessment by algorithm developers for the Square Kilometre Array (SKA; \citealt{SKA1}). In contrast, our method adopts a different strategy, evaluating image quality by analysing the entire image without the need for source extraction. Notably, to compare the object (no matter source or noise) at the same position in the two input images, we assign greater significance to sources based on their intensities, all without relying on empirical thresholds, while also considering background noise.

\section{Radio interferometric imaging of extended sources}
\label{section2}

Radio emissions of celestial sources in the sky as detected by an interferometer are known as visibilities \citep{RAbook}. In an interferometer, the vector that connects the phase centres of two antennas is called a baseline vector. The relationship between observed complex visibilities and sky brightness distribution (SBD) associated with the baseline vector is a Fourier transform under certain conditions \citep{RAbook} and can be described (\citealt{RA7}) by
\begin{equation}
V\left( {u,v,w} \right) = {\iint{\frac{B\left( {l,m} \right)}{\sqrt{1 - l^{2} - m^{2}}}e^{- i2\pi{({ul + vm + w{({\sqrt{1 - l^{2} - m^{2}} - 1})}})}}dldm}} ,
\label{vis}
\end{equation}
where \textit{V} is the observed complex visibility, \textit{B} is the distribution of sky brightness, \textit{l,m,n} are coordinates in the spatial domain, and \textit{u,v,w} are coordinates in the frequency domain.

The inversion from visibilities to image is an ill-posed problem due to limited number of antennas \citep{RA3,RA4}. Consequently, Radio interferometric (RI) imaging methods are developed to restore the SBD. The most pragmatic RI imaging method is CLEAN \citep{CLEAN1}, which is a non-linear deconvolution algorithm. The most traditional CLEAN \citep{CLEAN1} method is an iterative process used to extract point sources. However, estimating the sky brightness by point sources is not a sensible assumption for extended sources. There are several RI imaging methods that are suitable for extended source imaging, such as multi-scale (MS) CLEAN \citep{CLEAN2}, the Maximum Entropy Method (MEM; \citealt{MEM10}), and compressed sensing (CS) based methods, e.g., Isotropic Undecimated Wavelet Transform (IUWT; \citealt{iuwt, CS16, CS14, RA3}).

\subsection{MS-CLEAN}
\label{msclean}

MS-CLEAN has been proposed \citep{CLEAN2} to deal with the extended sources by multi-scale decomposition. MS-CLEAN looks for the maximum peak of all residual images with different scales and then adds the peak to the model image with a certain loop gain. When the noise level of the residual image is under a certain criteria, a model image is generated, estimating the true sky brightness.

There are many data processing software packages that can be used to implement RI imaging. The most well-known are Common Astronomy Software Applications (CASA; \citealt{CASA,CASAnew}), W-Stacking Clean (WSClean; \citealt{WS1,WS2,WS3}), and Radio Astronomy Simulation, Calibration and Imaging Library (RASCIL) \footnote{\url{https://developer.skao.int/projects/rascil/en/latest/}}. They all have functionality to implement MS-CLEAN. Herein, we use CASA and WSClean to implement MS-CLEAN.

\subsection{MEM}

Another RI imaging method that is suitable for imaging extended sources is MEM \citep{MEM2,MEM7,MEM8,MEM9}. A function $F\left( I_{M} \right)$ is defined as the entropy of the intensity distribution of the model $I_{M}$. One of the best entropy functions \citep{MEM} is 
\begin{equation}
    F = - {\sum\limits_{k}{I_{M_k}ln\frac{I_{M_k}}{I_{M^{0}_{k}}e}}} ,
    \label{mem1}
\end{equation}
where $I_{M}$ is the model image and $I_{M^{0}}$ is the priori model image. The sum is taken over every pixel $k$ in the image. In the imaging process, the entropy is maximised subject to the consistency between the model visibility and the measured visibility. The solution is an iterative process \citep{MEM10,MEM2,MEM3,MEM6}.

Herein, we use GPUVMEM \citep{GPUVMEM}, which is a GPU-accelerated MEM software package, to perform MEM in Section 4.

\subsection{IUWT}

IUWT is a compressed sensing (CS) based RI imaging method which has good performance for imaging extended sources. CS is a robust technique for signal reconstruction. For a sparse image, the image size can be compressed far below the requirement of Nyquist sampling \citep{RAbook2}. CS fosters sparse regularisation methods that can be used to address inverse problems, such as the inverse transformation from visibility to image in RI imaging \citep{iuwt}. Formal requirements (e.g., sparsity and incoherent sampling) of CS can be roughly met in RI imaging \citep{RAbook2}. The sensing matrix indicates the mathematical relationship between the image to be reconstructed and the measured visibilities. If the sensing matrix obeys the Restricted Isometry Property (RIP; \citealt{RIP,RA6}), CS is a promising method for reconstructing the SBD \citep{BigData11}.

Alternative CS methods have been applied to aperture synthesis \citep{CS14,CS15,CS19}, e.g., PURIFY \citep{RA6}, MOdel REconstruction by Synthesis-ANalysis Estimators (MORESANE; \citealt{RA3}) and SparseRI \citep{CS18}, rotation measure synthesis \citep{CS16,CS17}, and CS-ROMER \citep{csromer}. Herein, the reweighted version \citep{iuwt_code} of IUWT is used in Section 4.

\section{Image quality assessment methods}

In this section, we start by reviewing a well-known IQA method, the SSIM index, and two IQA methods proposed in \citealt{Wider}, the ITW-SSIMs with different weighting functions and LISI. Herein, we supplement the ITW-SSIMs and LISI with more details of the derivation and maths. ITW-SSIMs and LISI aim to assess low-information images, for example, RA images with point sources. However, RA images with extended sources include complicated structure in the non-residual part. Therefore, in this article, we propose a new variant of LISI, augmented LISI (augLISI), to assess images containing extended sources in RI imaging. ITW-SSIM, LISI, and augLISI are open source available on our GitHub page\footnote{\url{https://github.com/egbdfX/Intensity-sensitive-IQAs}} and Zenodo \citep{zenodo}.

\subsection{SSIM}
The equation describing SSIM ($\mathrm{SSIM} \in \left\lbrack - 1,1 \right\rbrack$) is given by
\begin{equation}
\label{SSIM}
    \mathrm{SSIM}\left( {p,q} \right) = \frac{\left( {2\mu_{p}\mu_{q} + C_{1}} \right)\left( {2\sigma_{pq} + C_{2}} \right)}{\left( {\mu_{p}^{2} + \mu_{q}^{2} + C_{1}} \right)\left( {\sigma_{p}^{2} + \sigma_{q}^{2} + C_{2}} \right)},
\end{equation}
where $\mu_{p}$ (or $\mu_{q}$) is the mean intensity of image \textbf{\textit{p}} (or \textbf{\textit{q}}), $\sigma_{p}$ (or $\sigma_{q}$) is the standard deviation of the pixel intensities in image \textbf{\textit{p}} (or \textbf{\textit{q}}), $\sigma_{pq}$ is the cross covariance of pixel intensities in image \textbf{\textit{p}} and image \textbf{\textit{q}}, and $C_{1}$ and $C_{2}$ are constant values to avoid instability when the denominator is very small.

Currently, SSIM is widely discussed and used in IQA. However, the result of SSIM is uninformative for assessing low-information images \citep{Wider}, such as RA images, where the importance of different objects in the images is different. Celestial sources are the most significant information in RA images. The SSIM results for RA images always approach 1, due to the limited number of sources and the large noise-like regions in the images. Admittedly, some state-of-the-art SSIM also weight the importance of different information, such as multi-scale SSIM \citep{SSIM4}, G-SSIM \citep{SSIM5}, and IW-SSIM \citep{SSIM10}. However, they are not suitable for assessing RA images, according to the experimental results in \citealt{Wider}.

Therefore, ITW-SSIMs and LISI have been developed in \citealt{Wider} for low-information RA images.

\subsection{IQA for images that only contain point source(s)}

In ITW-SSIMs \citep{Wider}, different weighting functions are applied to weight the importance of each pixel according to its intensity. The mathematical derivation of it is provided in Appendix \ref{itwsection}. In addition to ITW-SSIMs, LISI is another IQA index developed to assess low-information images. The amount of information contained in the inputs should be reflected in an effective intensity-sensitive IQA index. The similarity indicated by the index should be higher when the pixels at the same position in the inputs have similar intensities and the intensities are both high. Following this principle, LISI \citep{Wider} (whose range should be $\left\lbrack 0,1 \right\rbrack$) is expressed as 
\begin{equation}
\label{LISI}
\mathrm{LISI}\left( {x,y} \right) = D\frac{\sum\limits_{i=1}^{N}\frac{\left| {x_{i} + y_{i}} \right|}{\left| {x_{i} - y_{i}} \right| + C_{1}}}{\max\left( {\sum\limits_{i=1}^{N}x_{i}},{\sum\limits_{i=1}^{N}y_{i}} \right) + C_{2}},
\end{equation}
where $C_{1}<<1$, $C_{2}<<1$, and the inputs $x$ and $y$ are normalised.

To set the factor $D$ in a meaningful way, LISI is supposed to be 1 when the two input images are the same ($x_{i} = y_{i}$ so that $\left| {x_{i} - y_{i}} \right|=0$) and informative (with high intensity parts), which can be expressed as
\begin{equation}
\label{D1}
    \mathrm{LISI}\left( {x,y} \right) = D\frac{\sum\limits_{i=1}^{N}\frac{\left| {x_{i} + y_{i}} \right|}{\left| {x_{i} - y_{i}} \right| + C_{1}}}{\max\left( {\sum\limits_{i=1}^{N}x_{i}},{\sum\limits_{i=1}^{N}y_{i}} \right) + C_{2}} = D\frac{\sum\limits_{i=1}^{N}\frac{{2x}_{i}}{0 + C_{1}}}{{\sum\limits_{i=1}^{N}x_{i}} + C_{2}} = \frac{2D}{C_{1}}\frac{\sum\limits_{i=1}^{N}x_{i}}{{\sum\limits_{i=1}^{N}x_{i}} + C_{2}} \approx \frac{2D}{C_{1}} = 1.
\end{equation}

LISI is supposed to be 0 when the two input images are entirely different, i.e., $\forall i, x_{i} \neq y_{i}$ and $\left| {x_{i} - y_{i}} \right|$ is very large (to determine D, we make an assumption that $\forall i, x_{i} > y_{i}$), which can be expressed as
\begin{equation}
\label{D2}
    \mathrm{LISI}\left( {x,y} \right) = D\frac{\sum\limits_{i=1}^{N}\frac{\left| {x_{i} + y_{i}} \right|}{\left| {x_{i} - y_{i}} \right| + C_{1}}}{\max\left( {\sum\limits_{i=1}^{N}x_{i}},{\sum\limits_{i=1}^{N}y_{i}} \right) + C_{2}} \approx D\frac{\sum\limits_{i=1}^{N}\frac{x_{i}}{x_{i} + C_{1}}}{{\sum\limits_{i=1}^{N}x_{i}} + C_{2}} \approx D\frac{\sum\limits_{i=1}^{N}1}{{\sum\limits_{i=1}^{N}x_{i}} + C_{2}} = D\frac{N}{N' + C_{2}} \approx 0,
\end{equation}
where the $N'$ is no greater than $N$.

Therefore, $D$ should be $D = \frac{C_{1}}{2}$ and very small.

\subsection{IQA for images that contain extended source(s)}
\label{Sec33}
One of the advantages of ITW-SSIMs and LISI is that they do not require an empirical threshold to determine whether a certain pixel belongs to noise or a celestial source. According to the expression of the numerator of LISI, besides large differences in very high-intensity parts, LISI is also sensitive to large differences in lower-intensity parts. When an image contains extended source(s), there will be more differences in the lower-intensity part due to the increased complexity of the image. In such cases, LISI may be overly sensitive.

\begin{figure*}
\gridline{\fig{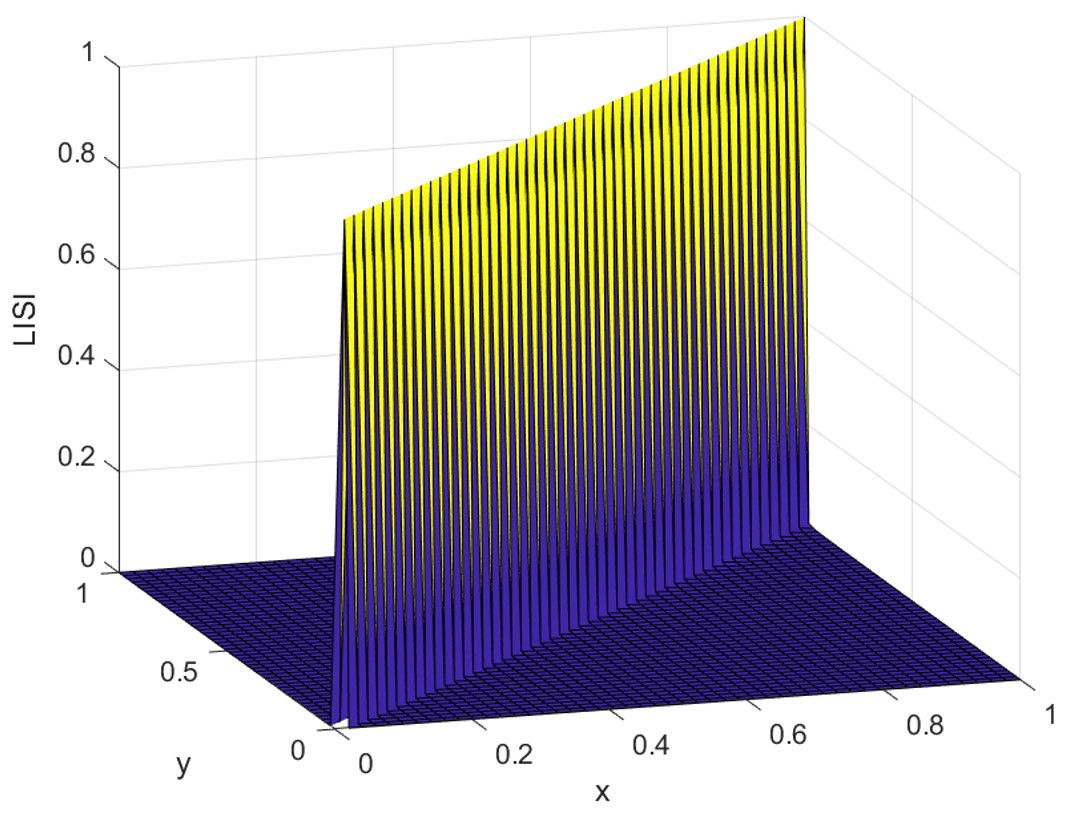}{2in}{(a)}
          \fig{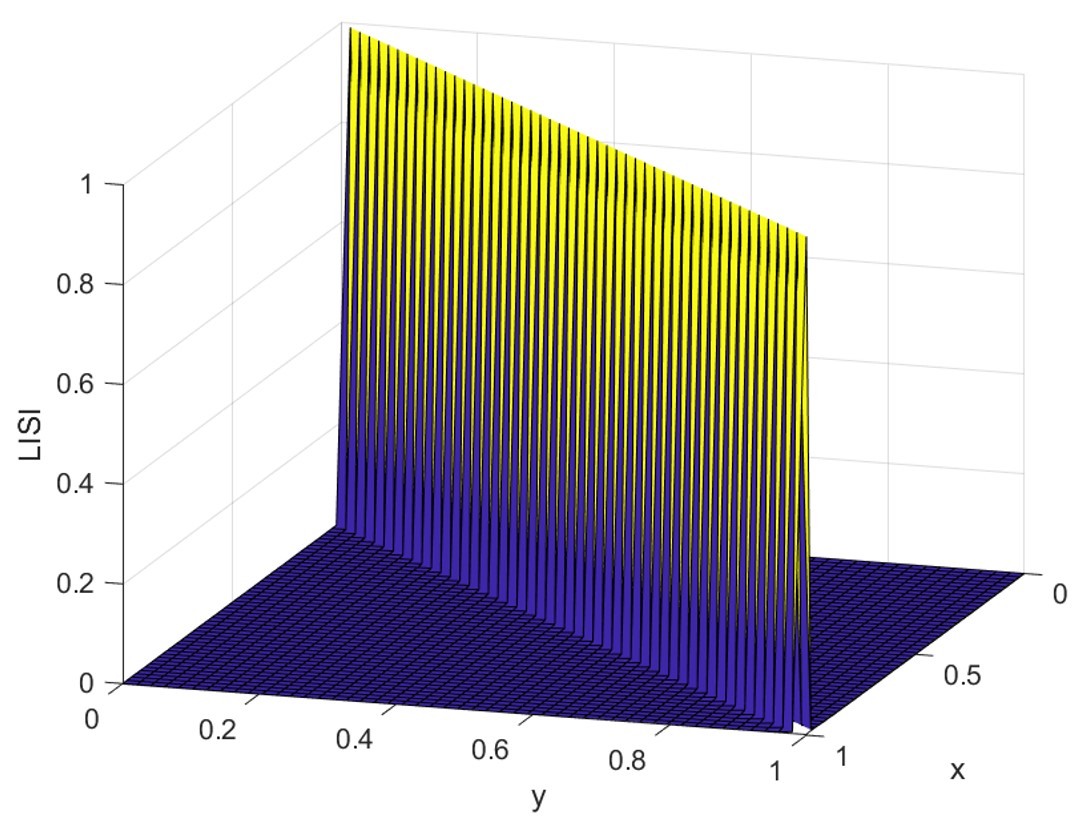}{2in}{(b)}
          \fig{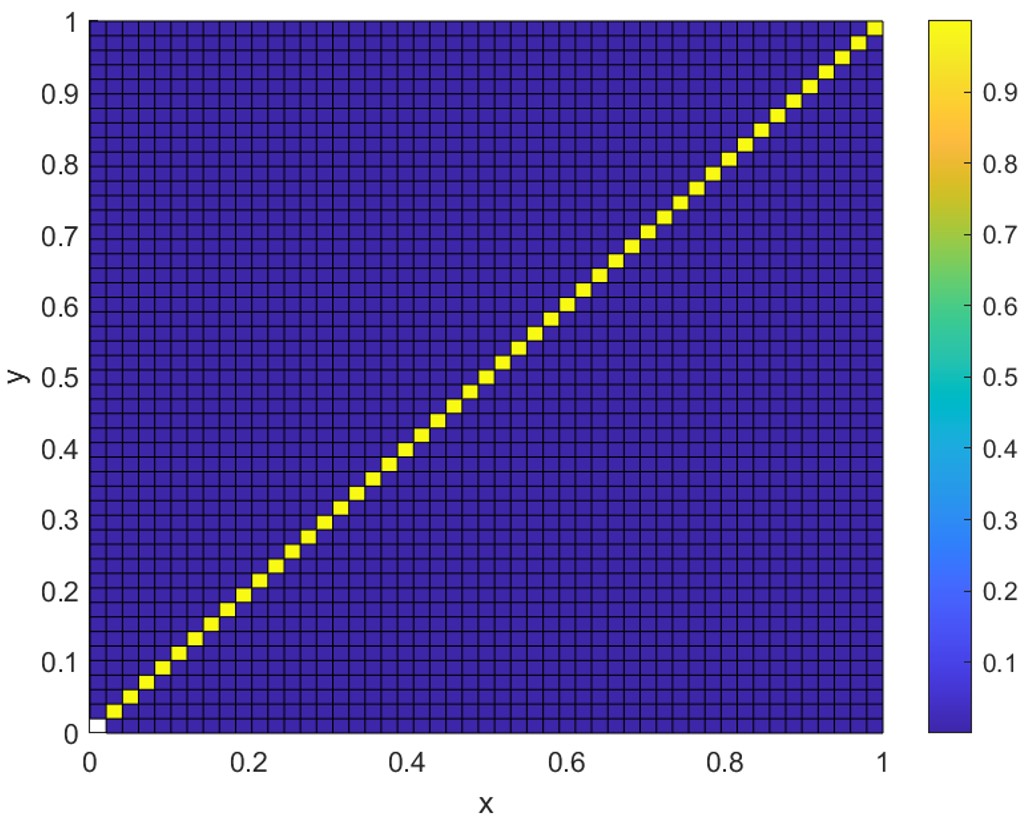}{2in}{(c)}
          }
\gridline{\fig{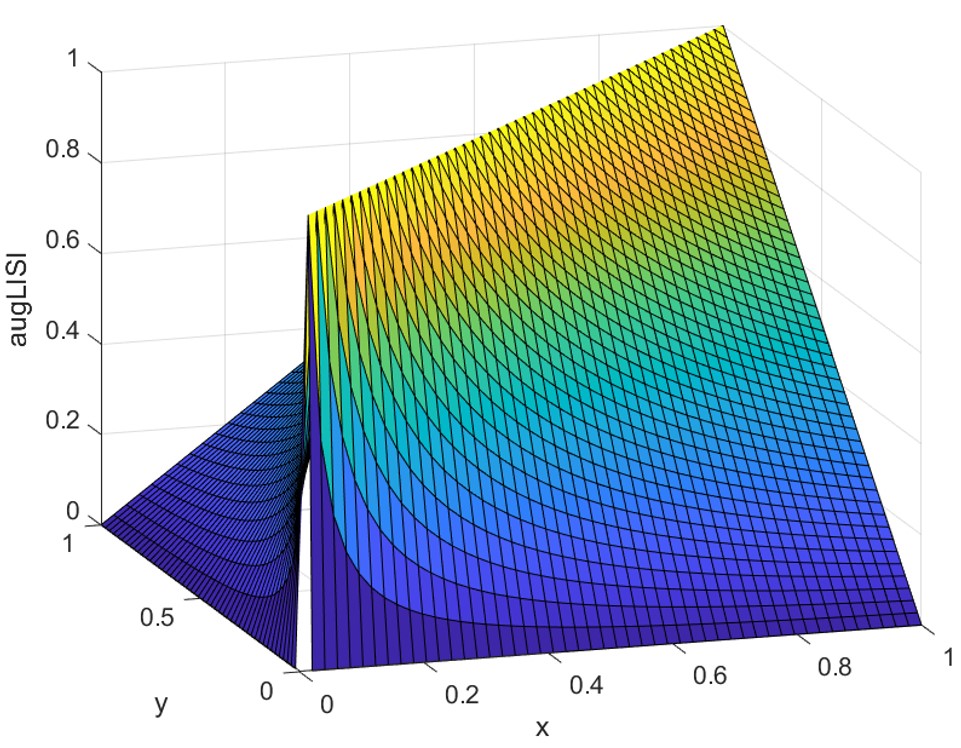}{2in}{(d)}
          \fig{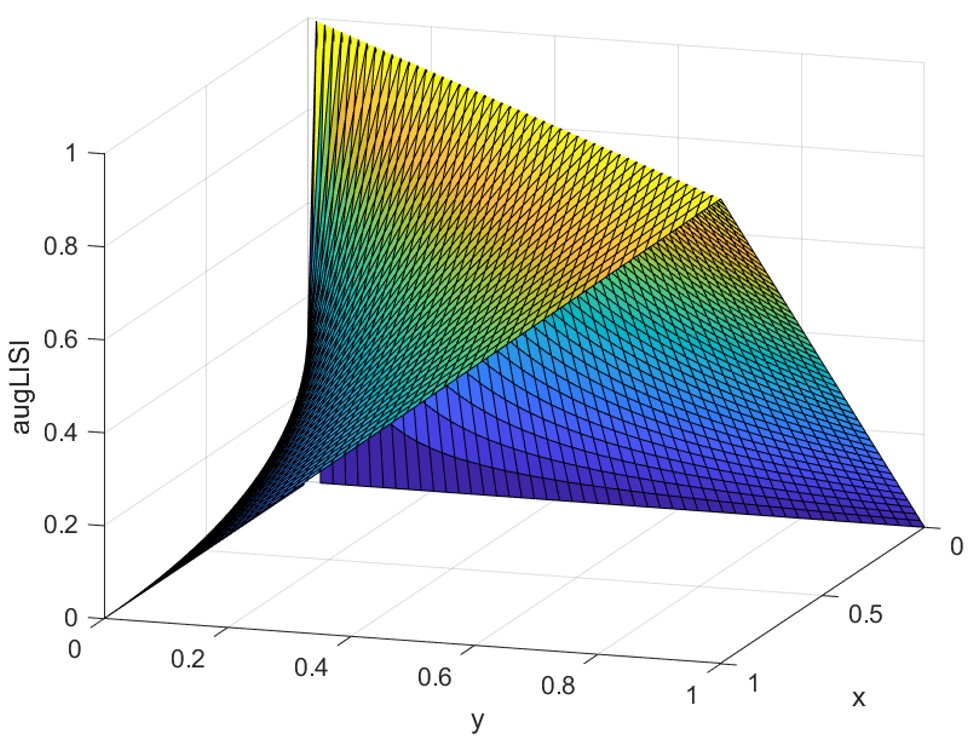}{2in}{(e)}
          \fig{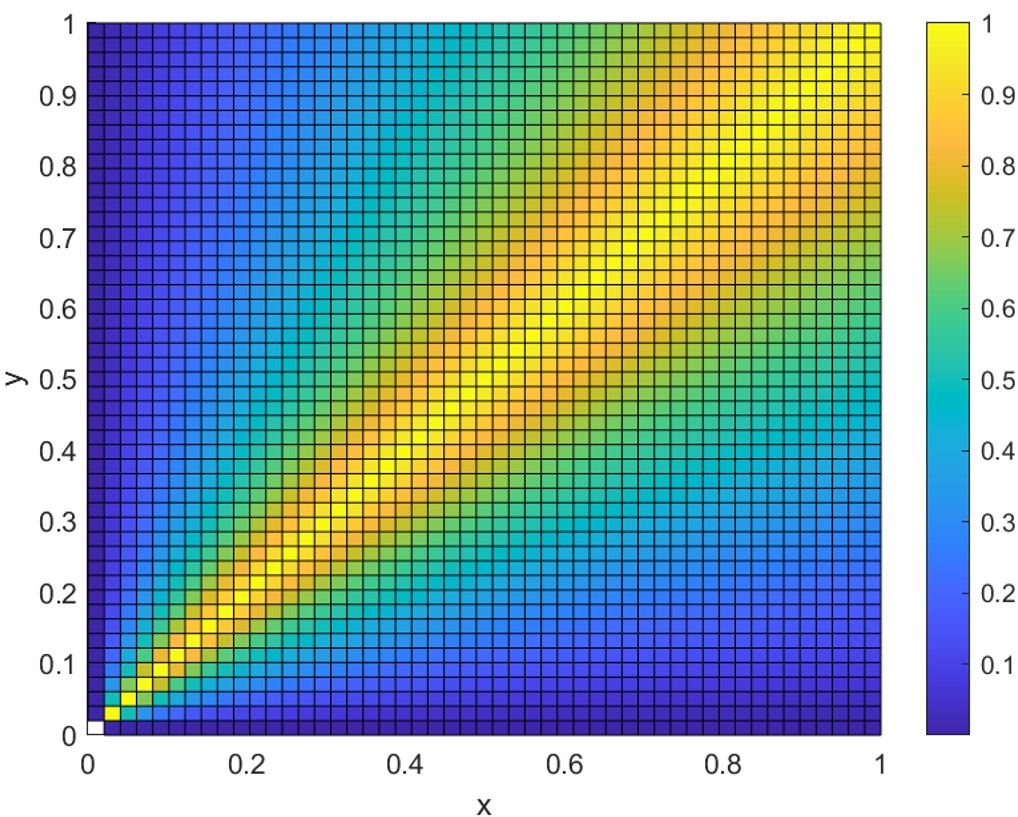}{2in}{(f)}
          }
\caption{Response of LISI (a, b, c) and augLISI (d, e, f) to a pair of single-pixel images $(x,y)$, where intensity $x \in [0,1]$ and $y \in [0,1]$. To show these 3D plot, three directions are illustrated, separately.\label{empiri}}
\end{figure*}

To illustrate this, a simulation can be performed based on Equation \ref{LISI}. The response of LISI to a pair of single-pixel images $x$ and $y$ is shown in Fig. \ref{empiri} (a) - (c), where both the pixel values of $x$ and $y$ are between 0 and 1. As we can see from the figure, LISI does emphasis the high-intensity similarity (see the diagonal portion of the figure) but when lower-intensity parts are extended and have differences, the value of LISI will decrease dramatically indicating that the differences mean a lot to the assessment; in other words, LISI is over-sensitive in this case.

To make the index only sensitive to large differences in celestial sources, a new variant of LISI, augmented LISI ($\mathrm{augLISI} \in \left\lbrack 0,1 \right\rbrack$), is proposed in this article. When two images are identical, $\mathrm{augLISI}$ will be 1.

It is expressed as
\begin{equation}
    \label{augLISI}
    \mathrm{augLISI}\left( {x,y} \right) = 1 - \frac{\sum\limits_{i=1}^{N}\left| {x_{i} + y_{i}} \right| \left| {x_{i} - y_{i}} \right|}{\sum\limits_{i=1}^{N}x_{i}+\sum\limits_{i=1}^{N}y_{i}+ C},
\end{equation} 
where $C<<1$ and the inputs $x$ and $y$ are normalised.

It is worth noting that in LISI and augLISI, the denominators are to reduce the effect of different sizes and intensity levels of the inputs.

The numerator of augLISI is designed to be the multiplication of the sum and the difference of the corresponding pixels in the two input images, rather than division as with LISI. This adjustment ensures that the IQA index is less sensitive when the two inputs are very different but both have low intensities, since this would indicate that the noise is very different. The denominator of augLISI is designed to be the sum of all pixels in the two input images, rather than using the maximum sum of the two input images. This adjustment is made to avoid the case where a bright image would dominate a dark image in the comparison, a situation that can occur, for example, when assessing the output from different iterations of iterative algorithms used in radio astronomy. The response of augLISI is shown in Fig. \ref{empiri} (d) - (f), which emphases high-intensity differences, but becomes smoother for lower-intensity differences.

\begin{figure*}[!t]
\gridline{\fig{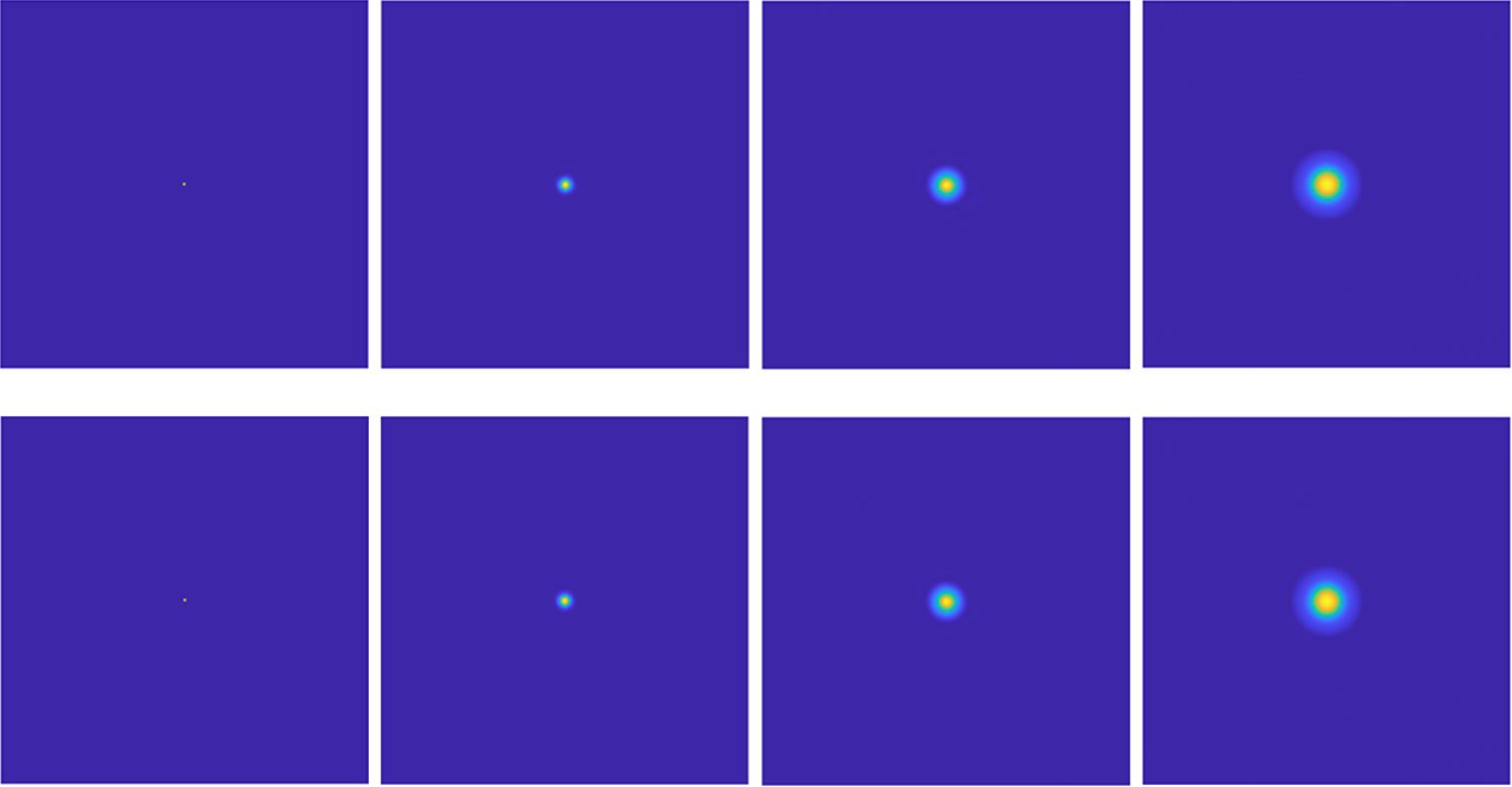}{4in}{(a)}
          \fig{sources.jpg}{3in}{(b)}
          }
\caption{Simulated Gaussian sources with different FWHMs of major and minor axes, where (a) shows four pairs (vertical) of examples, with $\Gamma = 0, 13, 26, 39$ from left to right, and (b) shows a sketch of the two sources in a pair.
\label{gauimages}}
\end{figure*}

\begin{figure}[!t]
	\centering
\includegraphics[width=3in]{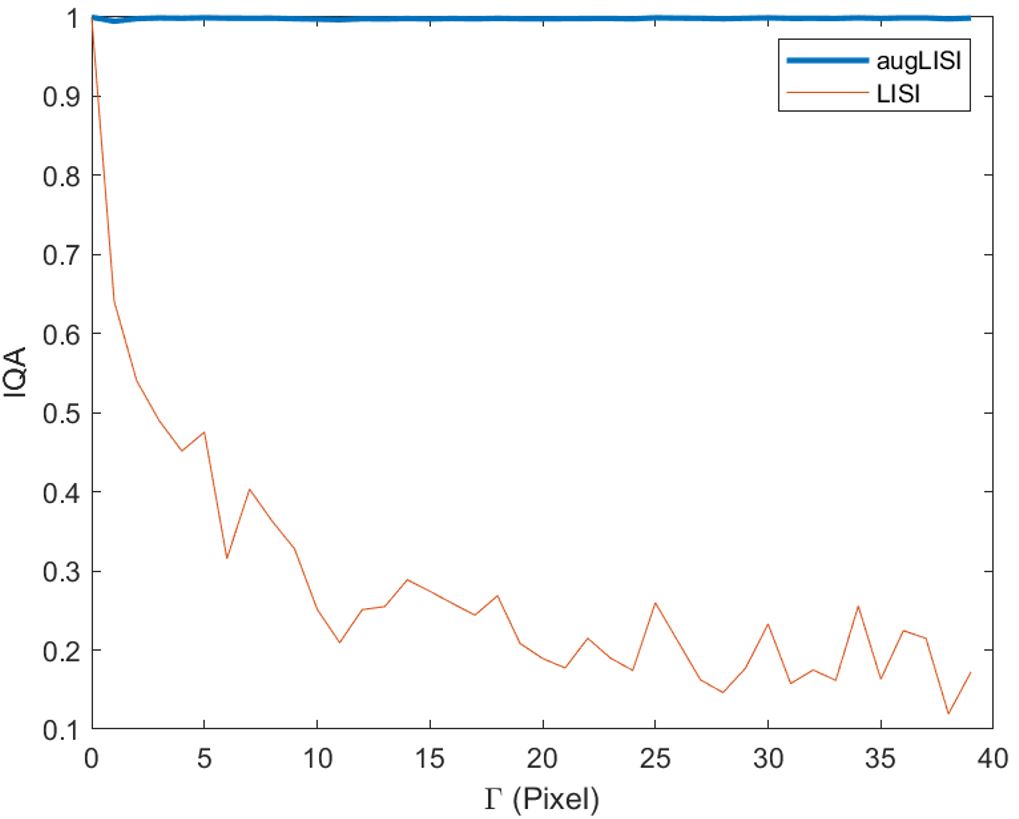}
\caption{LISI and augLISI results on pairs of images with different sizes of simulated Gaussian sources.
\label{gausimulation}}
\end{figure}

Along with the single-pixel simulation, an RA-style data simulation has been done to further support the argument. Oxford's Square Kilometre Array Radio-telescope simulator (OSKAR; \citealt{OSKAR,OSKAR1}) is used to simulate a two-dimensional elliptical Gaussian source having different sizes. In the simulation, 40 pairs of images are generated using the SKA AA2 telescope array layout \citep{aa2,aa22}, where in each image there is a single source at the centre. The size of source is defined by the full width at half maximum (FWHM) $\Gamma$ ($\Gamma = 0,1,2,...,39$ in units of pixels) of the major and minor axes. In each pair of images, the source in one of the images has major and minor axes equal to $\Gamma$; the source in the other image has major and minor axes equal to $\Gamma+0.1$ and $\Gamma$, respectively. Note that `pixel' is used as the unit of source size, rather than using angular resolution, for the reason that the IQA matrix is solely based on input images, which should not be affected by the resolution of telescopes. Examples of the simulated Gaussian sources are shown in Fig. \ref{gauimages}. When using IQA metrics to compare each pair of images with the same $\Gamma$ value for the minor axis, the difference between the two images in each pair is expected to be small, and the IQA should be stable (nearly constant) with the increase in source size, as the difference of 0.1 is negligible when comparing the sizes of the sources. The IQA results are shown in Fig. \ref{gausimulation}. According to the result, LISI is more sensitive to the difference in sources, however it is over-sensitive to the difference especially when the size of source becomes larger.

LISI is sensitive to large differences in both high- and low-intensity parts of a structure, but will indicate that the inputs are similar only when the high-intensity part of the inputs have small differences. In contrast, augLISI is only sensitive to large differences in the high-intensity parts and will indicate that the inputs are similar in all the other cases.

\section{Performance on data from extended sources}

In this section, we demonstrate the application of our IQA methods for RI imaging by using two real datasets. The tutorial data from \citep{extended_data} is data measured by the Karl G. Jansky Very Large Array (VLA) of the extended source, supernova remnant G055.7+3.4 (referred to as `SNR\_G55\_10s' hereafter). Along with this, another tutorial data from \citep{smiledata} is data measured by VLA of the extended source, binary black hole system 3C 75 in the Abell 400 cluster of galaxies (referred to as `3C75' hereafter). Section \ref{41} - \ref{44} demonstrate contributions 1-3 presented in Section \ref{section1}, respectively. It is worth noting that although we used the SKA AA2 telescope array layout to simulate data in Section \ref{Sec33}, we use datasets from the VLA telescope in this section to demonstrate that augLISI has no bias towards different telescope designs.

\subsection{Comparing different implementation approaches for the same imaging algorithm}
\label{41}

The development of different imaging algorithms and codes are typically undertaken by disparate research teams often associated with different instruments. Teams formulate distinct implementation approaches for a given algorithm, grounded in varying development methodologies. For example, both CASA and WSClean have implementations of MS-CLEAN. However, they differ in their implementation, with the most significant difference between them being the process of undertaking gridding. Specifically, WSClean adopts \textit{w}-stacking \citep{w5,WS1,w7} for this purpose, while CASA does not incorporate this approach. CASA proffers an array of alternatives for gridding, encompassing standard (prolate spheroid), \textit{w}-projection, facetted, \textit{a}-projection, and \textit{aw}-projection (for more information, refer to ``tclean'' in CASA \citep{casaweb}).

\begin{figure*}
\gridline{\fig{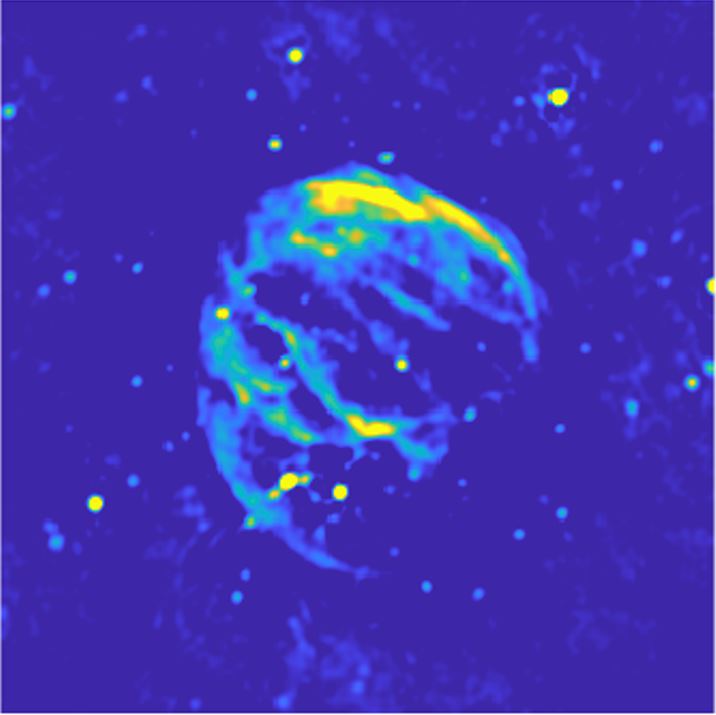}{2in}{(a)}
          \fig{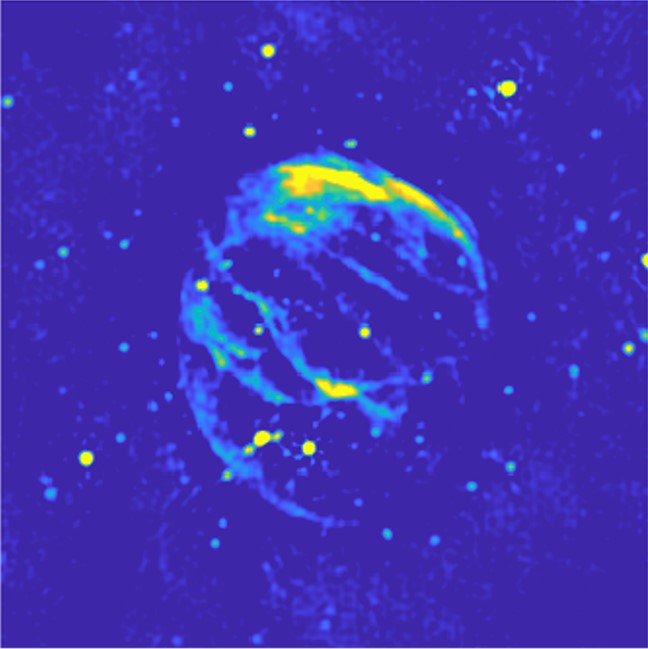}{2in}{(b)}
          }
\gridline{\fig{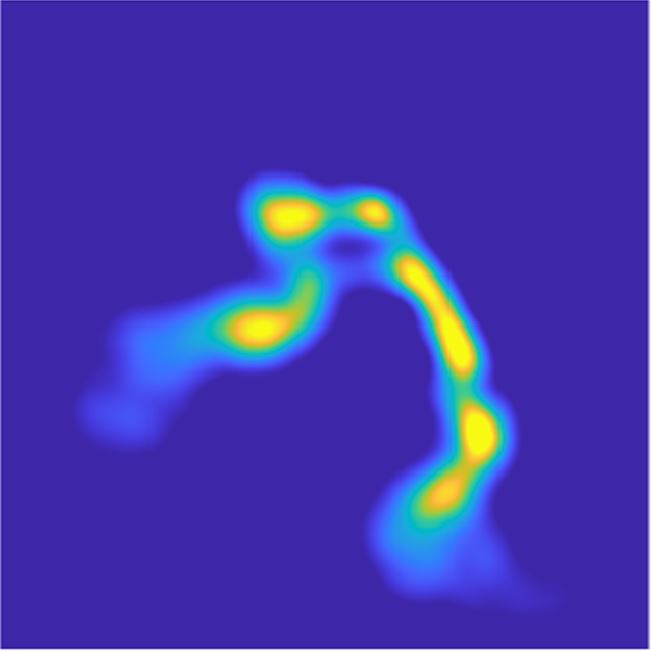}{2in}{(c)}
          \fig{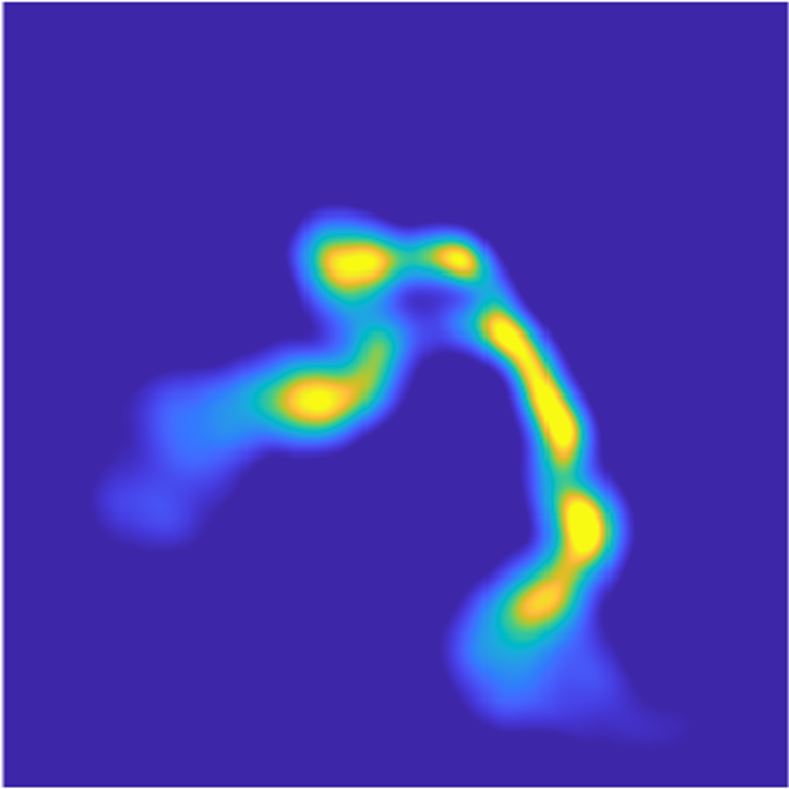}{2in}{(d)}
          }
\caption{MS-CLEAN results of SNR\_G55\_10s generated by (a) CASA and (b) WSClean, and MS-CLEAN results of 3C75 generated by (c) CASA and (d) WSClean.
\label{ms}}
\end{figure*}

Figure \ref{ms} shows the MS-CLEAN results for SNR\_G55\_10s and 3C75 generated by CASA and WSClean. To make the outputs comparable, for the same dataset, both implementations use the same image size, cell size, size of CLEAN beam, weighting mode (Briggs weighting; \citealt{briggs}), and maximum number of iterations. As WSClean adopts \textit{w}-stacking which is suitable for wide-field imaging, we set CASA to run \textit{w}-projection to make them as comparable as possible.

\begin{figure}[!t]
    \centering
    \includegraphics[width=5in]{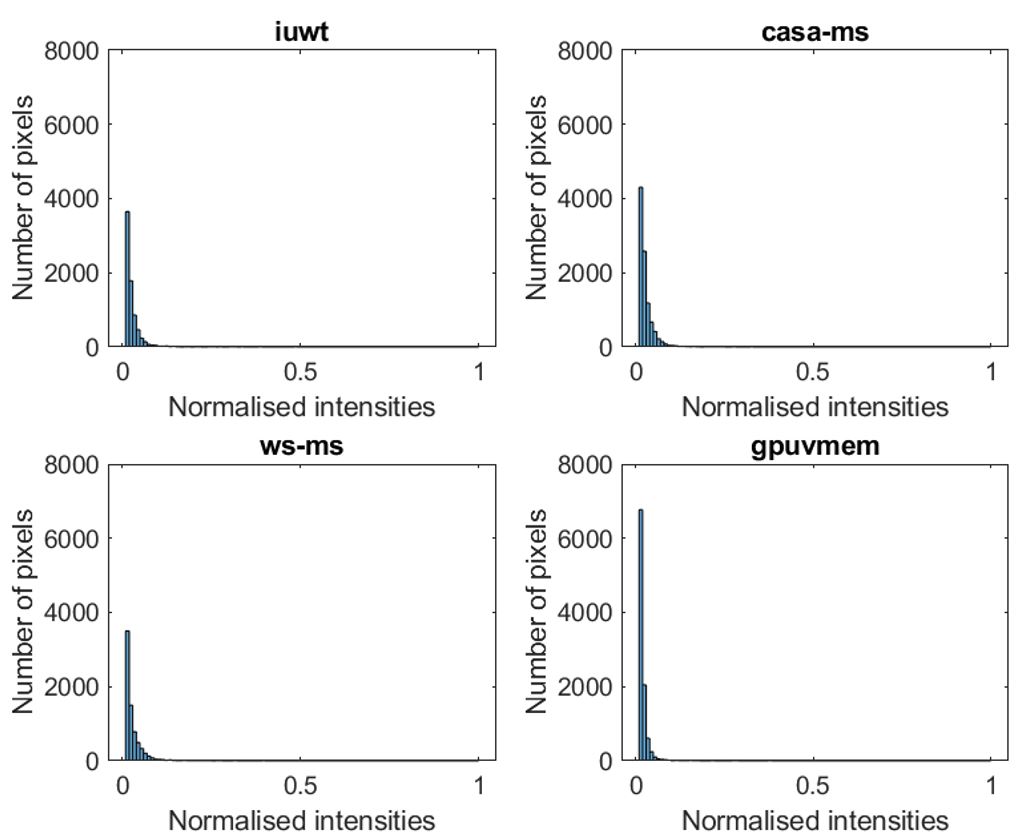}
\caption{Histograms of SNR\_G55\_10s for the results of (top-left) IUWT, (top-right) MS-CLEAN by CASA, (bottom-left) MS-CLEAN by WSClean, and (bottom-right) MEM by GPUVMEM. The distributions are depicted by histograms of the normalised intensities of each pixel. To show the histograms in meaningful ways, all 0 values are ignored when plotting the histograms.}
\label{noise}
\end{figure}

\begin{figure}[!t]
    \centering
    \includegraphics[width=5in]{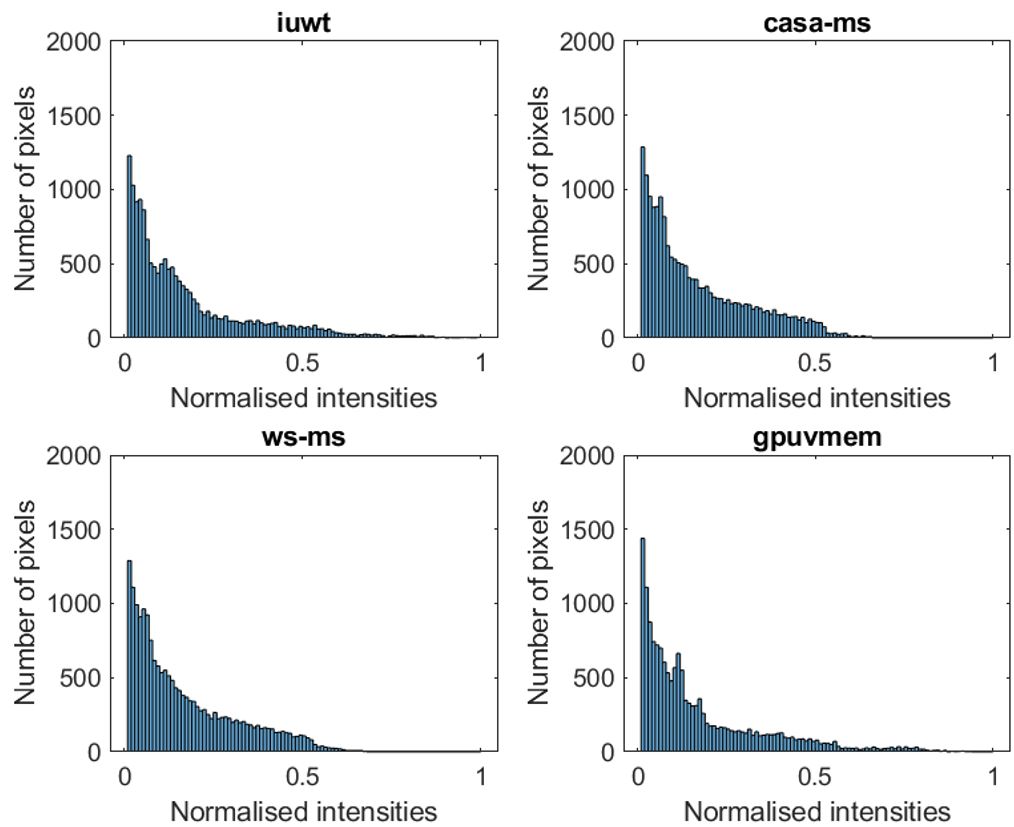}
\caption{Histograms of 3C75 for the results of (top-left) IUWT, (top-right) MS-CLEAN by CASA, (bottom-left) MS-CLEAN by WSClean, and (bottom-right) MEM by GPUVMEM. The distributions are depicted by histograms of the normalised intensities of each pixel. To show the histograms in meaningful ways, all 0 values are ignored when plotting the histograms.}
\label{noise-cas}
\end{figure}

It is worth noting that normalisation needs to be done in pre-processing of IQA to ensure the intensity levels of the images are comparable. In the pre-processing, we firstly calculate the z-score ($(image - \mathrm{mean}(image))/\mathrm{std}(image)$, where $\mathrm{std}$ means standard deviation) of each image, and then normalise it with the maximum pixel value among all the four images (i.e., images generated by MS-CLEAN in CASA, MS-CLEAN in WSClean, GPUVMEM, and IUWT) so that pixel values of each image are between 0 and 1. The histograms of the images after pre-processing are shown in Fig. \ref{noise} for SNR\_G55\_10s and in Fig. \ref{noise-cas} for 3C75.

The comparisons using SSIM and augLISI are presented in Table \ref{ms_iqa}, where images generated by MS-CLEAN in CASA and WSClean are compared for each dataset.
\begin{deluxetable*}{ccccc}
\tablecaption{SSIM and augLISI for assessing the results of different implementations of MS-CLEAN on real and simulated datasets.\label{ms_iqa}}
\tablewidth{0pt}
\tablehead{
\colhead{MS-CLEAN} & \colhead{Dataset} & \colhead{SSIM} & \colhead{augLISI}
}
\startdata
CASA vs WSClean & SNR\_G55\_10s (real) & 0.9846 & 0.9916\\
 & 3C75 (real) & 0.9983 & 0.9928\\
 & SNR\_G55\_10s (simulated) & 0.9793 ($\pm~0.0019$) & 0.9902 ($\pm~0.0005$)\\
 & 3C75 (simulated) & 0.9983 ($\pm~2.653\times {10}^{-5}$) & 0.9928 ($\pm~6.158\times {10}^{-5}$)\\
\enddata
\end{deluxetable*}

The results of SSIM (0.9846 for SNR\_G55\_10s and 0.9983 for 3C75, recall a value of 1 indicates the two images are identical) show that the different implementation approaches obtain similar results. SSIM not only reflects the similarity in high-intensity regions (likely to be source), but also reflects the difference in noise pattern. 

In contrast, augLISI only reflects the similarity in the high-intensity regions. The results obtained from augLISI confirm that CASA and WSClean produce very similar source in the output for SNR\_G55\_10s (0.9916, recall a value of 1 indicates the two images are identical) when using MS-CLEAN, despite the difference in implementation. By visual inspection, the output of CASA is more blurred than the output of WSClean. Since the value of augLISI is larger than that of SSIM, we conclude that the sources in the two images are basically the same and the difference is mainly due to noise or low-intensity structure. For 3C75, the value of augLISI is 0.9928, which is slightly smaller than SSIM (0.9983). But, importantly, the values of both augLISI and SSIM approach 1. In other words, the computational implementation of the same algorithm by two different groups using different programming languages and software engineering practices produces results that generally agree. 

Quantitatively, LISI, 0.0479 for SNR\_G55\_10s and 0.0491 for 3C75 (recall a value of 1 indicates the two images are identical), is over-sensitive for extended structures, as explained in Section \ref{Sec33}.

To illustrate the uncertainty of augLISI and SSIM in Table \ref{ms_iqa} in the presence of noise, we simulate datasets with a similar noise level as the real datasets but with different noise realisations. The noise level in the real dataset is estimated by fitting the histograms of the real and the imaginary parts of the visibilities, which are dominated by noise, with Gaussian functions using MATLAB's Open Curve Fitter app. For SNR\_G55\_10s, the R-square values when fitting to the real and imaginary parts are 0.9967 and 0.9968, respectively; for 3C75, the R-square values when fitting to the real and imaginary parts are 0.9623 and 0.9686, respectively. Different noise realisations are produced by generating different random values drawn from the relevant Gaussian distribution and adding these values back into the original dataset. This increases the noise in the original data but ensures that the level of noise in each noise realisation is the same. The results of SSIM and augLISI comparing the restored images generated by CASA and WSClean to the simulated datasets are shown in Table \ref{ms_iqa}. The mean values of augLISI and SSIM for the 10 different noise realisations are shown respectively in the table along with their uncertainties expressed in standard deviations.

We can see from Table \ref{ms_iqa} that augLISI is more stable than SSIM in the presence of noise. The IQA results of simulated datasets further support the above analysis of the real datasets. With augLISI, users are able to scientifically prove the similarity between different implementation approaches for the same algorithm, rather than judging intuitively. It is worth noting that SSIM and augLISI have similar performance in the above examples, because both noise pattern and source structure are similar in the two implementation approaches.

\begin{figure}[!t]
    \centering
    \includegraphics[width=7in]{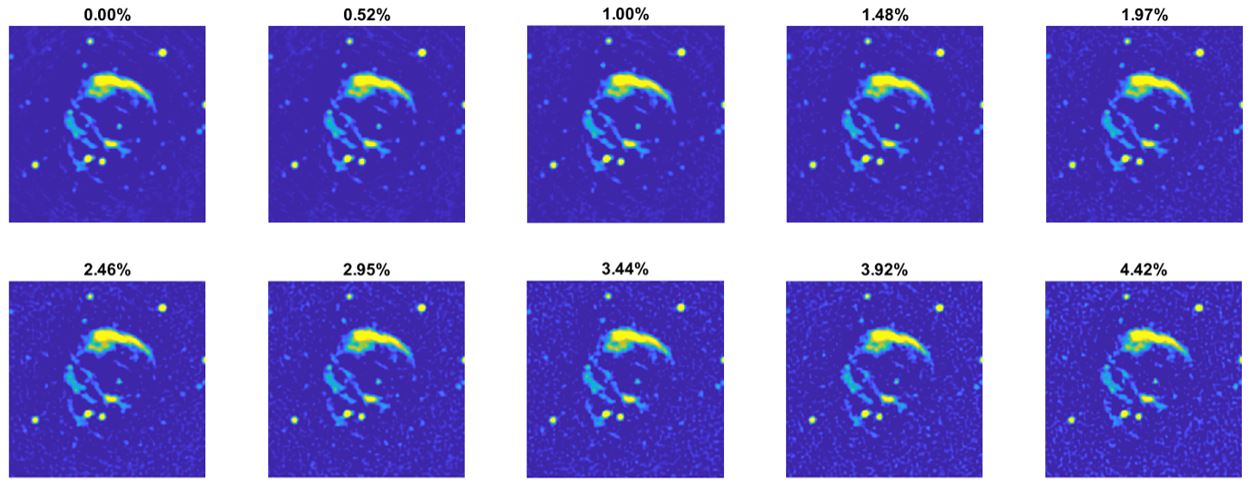}
\caption{Restored images of simulated SNR\_G55\_10s generated by WSClean. The 3-sigma additive noise levels of these figures are set to be a certain percentage (as shown above in each image) of the maximum intensity in the 0-additive-noise celestial source image.}
\label{noise41}
\end{figure}

To further demonstrate the enhancement of augLISI compared to SSIM and their uncertainties in the presence of noise, we simulate Measurement Sets with different noise levels. In OSKAR, noise is added to the visibilities by generating random numbers from a Gaussian distribution with a width given by the specified root mean square (RMS) value. Those values are then added independently to the real and imaginary components of the visibilities. This method is essentially the same as the one used in the simulation of Table 1. For each noise level, multiple noise realisations are generated by using different random seeds, these noise realisations are then used to statistically assess uncertainties. Take SNR\_G55\_10s as an example. The simulated datasets adopt the same telescope array as its real Measurement Set. We generated 100 simulated images of SNR\_G55\_10s with ten different noise levels, each level having ten different noise realisations using OSKAR. In the images, the 3-sigma random additive noise levels range from 0\% to 4.42\% of the maximum intensity of the celestial source simulated with no additive noise (which we will refer to as the 0-additive-noise celestial source), in steps of roughly 0.5\%, resulting in ten noise levels. We obtain MS-CLEAN (Briggs weighting) results by using WSClean on each image. A set of images restored by WSClean with different noise levels and different noise realisations are shown in Fig. \ref{noise41} as an example.

As the difference in the level of noise grows between input images (with the sources in those input images remaining the same), the difference between augLISI and SSIM will grow. To illustrate this, Table \ref{table41} shows the SSIM and augLISI results comparing the 0-additive-noise image with increasingly noisy images, with the standard deviation (Std) included below the corresponding mean of the IQA. When the noise increases, we still expect the images to be assessed as ‘similar’ because the sources are same in these datasets. It can be seen from the results that SSIM is affected more by the noise than augLISI. The gap between the values of SSIM and augLISI becomes larger when the noise increases. Additionally, the standard deviations of SSIM are generally larger than those of augLISI, suggesting that augLISI increases the stability of assessing images in the presence of noise.
\begin{deluxetable*}{cccccccccccc}
\tablecaption{SSIM and augLISI of 0-additive-noise restored image and increasingly noisy restored images generated by simulated datasets.\label{table41}}
\tablewidth{0pt}
\tablehead{
\colhead{Noise level (\%)} & \colhead{}& \colhead{0 vs 0.00} & \colhead{0 vs 0.52} & \colhead{0 vs 1.00} & \colhead{0 vs 1.48} & \colhead{0 vs 1.97} & \colhead{0 vs 2.46} & \colhead{0 vs 2.95} & \colhead{0 vs 3.44} & \colhead{0 vs 3.92} & \colhead{0 vs 4.42}
}
\startdata
SSIM & Mean & 1.0000 & 0.9965 & 0.9874 & 0.9733 & 0.9553 & 0.9346 & 0.9120 & 0.8887 & 0.8647 & 0.8423\\
 & Std & 0.0000 & 0.0002 & 0.0005 & 0.0007 & 0.0011 & 0.0014 & 0.0017 & 0.0022 & 0.0024 & 0.0031\\
augLISI & Mean & 1.0000 & 0.9984 & 0.9972 & 0.9960 & 0.9947 & 0.9934 & 0.9921 & 0.9907 & 0.9894 & 0.9879\\
 & Std & 0.0000 & 0.0003 & 0.0002 & 0.0002 & 0.0002 & 0.0003 & 0.0003 & 0.0003 & 0.0004 & 0.0004\\
\enddata
\end{deluxetable*}

The values of SSIM and augLISI can be very different when the differences in noise pattern or source structure are larger. This will be further discussed in Section \ref{44}.

\subsection{Quantitatively defining the difference between different algorithms}
\label{42}

As CLEAN achieves extended imaging by applying the CLEAN beam, both MEM and compressed sensing methods are believed to be superior approaches for imaging extended sources. Our next research question pertains to the extent of improvement achieved by MEM and compressed sensing methods in comparison to CLEAN.

\begin{figure*}
\gridline{\fig{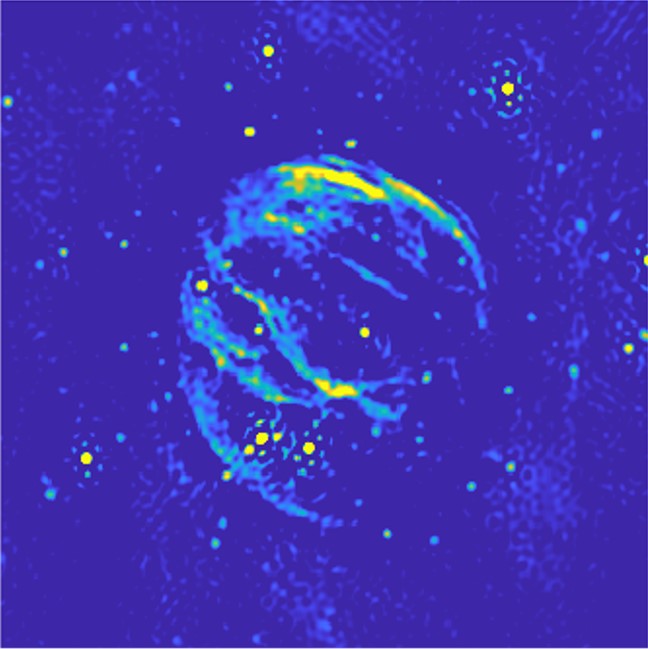}{2in}{(a)}
          \fig{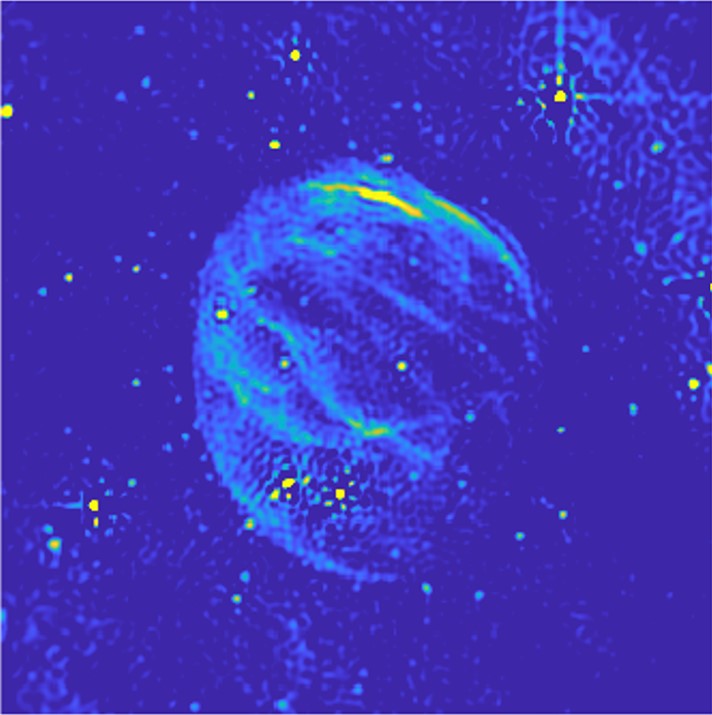}{2in}{(b)}
          \fig{ws-ms-zs.jpg}{2in}{(c)}
          }
\gridline{\fig{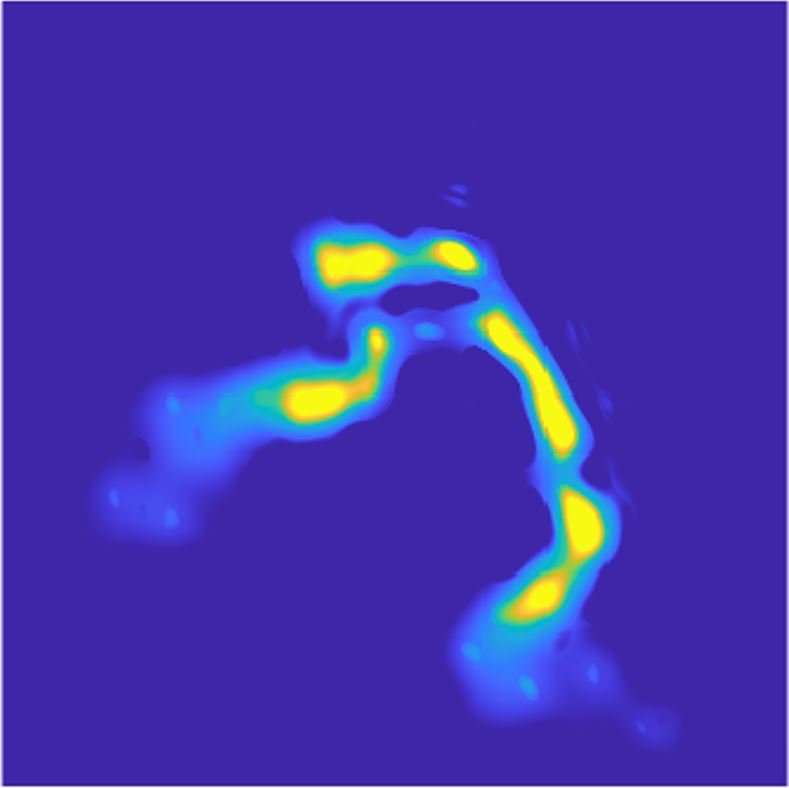}{2in}{(d)}
          \fig{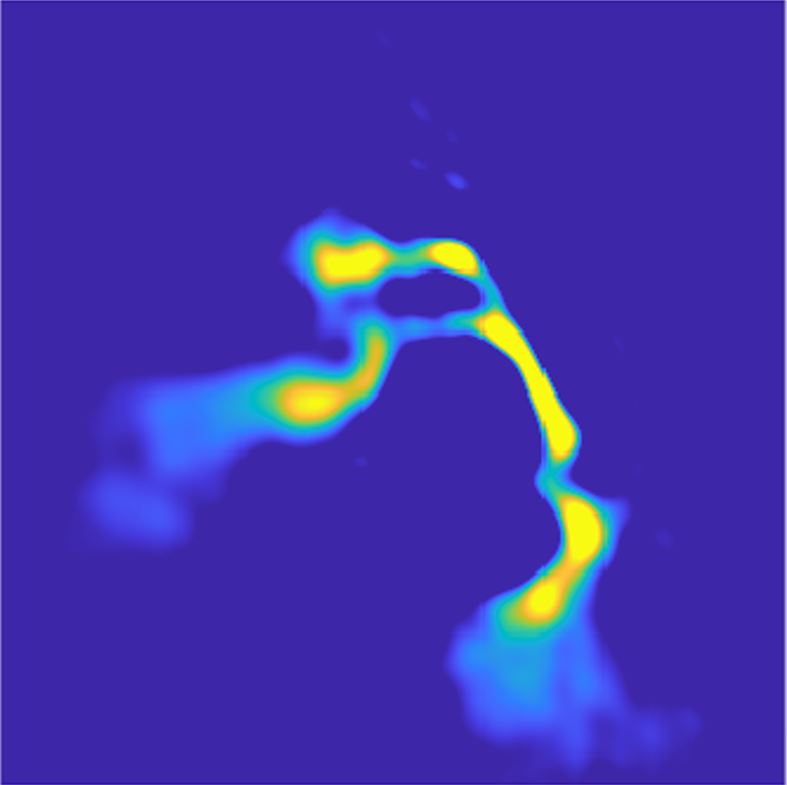}{2in}{(e)}
          \fig{ws-ms-zs-3c75.JPG}{2in}{(f)}
          }
\caption{(a) IUWT, (b) MEM, and (c) MS-CLEAN (WSClean, recalled Fig. \ref{ms}) results of SNR\_G55\_10s, and (d) IUWT, (e) MEM, and (f) MS-CLEAN (WSClean, recalled Fig. \ref{ms}) results of 3C75.
\label{iumem}}
\end{figure*}

Figure \ref{iumem} shows (a) the IUWT result and (b) the MEM result of SNR\_G55\_10s, and (d) the IUWT result and (e) the MEM result of 3C75. To make the images comparable, parameters are set in the implementations to be as close as possible with each other. In IUWT and GPUVMEM, the image size, cell size and weighting mode (Briggs weighting) are set by inputting either parameters or an example FITS file (e.g., the dirty image) which includes the information in the header. The termination condition in IUWT and GPUVMEM are different from the setting of CASA and WSClean. The process of IUWT terminates when the change of residual images between adjacent iterations is small enough, judging by Frobenius norm, as shown in
\begin{equation}
\epsilon = \frac{\| \Delta \mathrm{Residual} \|_{\textit{F}}^2}{\| \mathrm{Residual} \|_{\textit{F}}^2},
\end{equation}
where $\epsilon$ indicates the change, $\Delta \text{Residual}$ is the difference in residual images between adjacent iterations, and $\| \cdot \|_{\textit{F}}^2$ is the square of the Frobenius norm. The termination of GPUVMEM is based on the convergence of conjugate gradient \citep{GPUVMEM}. Although the setting modes are different in different implementations, we set them to be comparable based on the similarity of noise distribution (Fig. \ref{noise} and Fig. \ref{noise-cas}, these are the best cases in our experiments to make the noise as similar as possible).

Upon visual inspection, MEM provides better image quality for the extended source, comparing with the other three approaches. At the same time, IUWT on 3C75 obtains an image with a sharper boundary of the source. To quantitatively measure the extent of difference achieved by these algorithms, augLISI is used. Comparisons using SSIM and augLISI between MS-CLEAN (due to the similarity between the MS-CLEAN results generated by CASA and WSClean, we only consider the WSClean result for this comparison), IUWT, and MEM are presented in Table \ref{three_iqa}. We again see that SSIM (smaller than augLISI) reflects both the difference in residuals and high-intensity source structures in the images, while augLISI provides a more accurate comparison of the extended source restoration. Interestingly, in the comparison of MS-CLEAN and IUWT, for SNR\_G55\_10s, SSIM (0.9870) and augLISI (0.9870) are the same to four decimal places. Since SSIM does not have a preference about the difference in high or low intensity parts, but augLISI only focuses on high-intensity difference and is less sensitive to low-intensity difference. When SSIM and augLISI are same, it means that the difference between the two images concentrates on the high-intensity structure. This aligns with visual inspection.
\begin{deluxetable*}{cccc}
\tablecaption{SSIM and augLISI for assessing the results of different imaging algorithms\label{three_iqa}}
\tablewidth{0pt}
\tablehead{
\colhead{Dataset} & \colhead{Algorithms} & \colhead{SSIM} & \colhead{augLISI}
}
\startdata
SNR\_G55\_10s & MS-CLEAN vs MEM & 0.9554 & 0.9748\\
 & IUWT vs MEM & 0.9604 & 0.9767\\
 & MS-CLEAN vs IUWT & 0.9870 & 0.9870\\
3C75 & MS-CLEAN vs MEM & 0.9155 & 0.9277\\
 & IUWT vs MEM & 0.9326 & 0.9439\\
 & MS-CLEAN vs IUWT & 0.9435 & 0.9356\\
\enddata
\end{deluxetable*}

For the SNR\_G55\_10s data, the result of augLISI in Table \ref{three_iqa} shows that the similarity between IUWT and MS-CLEAN (0.9870) is higher than between MEM and either algorithm (0.9748 for MS-CLEAN and 0.9767 for IUWT). Upon visual inspection, the results of IUWT and MS-CLEAN are more similar, which confirms the augLISI results. Similarly, for the 3C75 data, generally the high-intensity parts of the source are successfully recovered by IUWT, MEM, and MS-CLEAN. Among the results of the three algorithms, MS-CLEAN produces the result with a diffuse boundary, while the boundary of the result generated by IUWT has a sharper transition. This leads to a smaller augLISI (0.9356) compared to SSIM (0.9435) when comparing the outputs of MS-CLEAN and IUWT. More details will be shown in the next Section for analysing the characteristics of the image using multiple IQA indexes jointly.

\subsection{Jointly analysing source and noise}
\label{44}

When IQA metrics indicate disparities, users may still find it challenging to discern whether these differences primarily arise from differences in the source(s) or in background noise. Observed differences in the source(s) could be resolved by adjustment of the imaging algorithm because the source(s) might not be fully resolved. If the difference is in the noise, we can anticipate that the imaging algorithm has restored the source(s) sufficiently. The issue can be more complicated for assessing an extended source(s), as differences can occur in any part of the extended structure.

Analysing signal and noise together can be useful in signal processing, for example, in acoustics \citep{acoustics} and auditory processing \citep{acoustics2}. For assessing RA images, researchers can analyse the noisy image by extracting faint parts \citep{faiextra} or extracting contours and textures \citep{contexextra}.

\begin{figure*}
\gridline{\fig{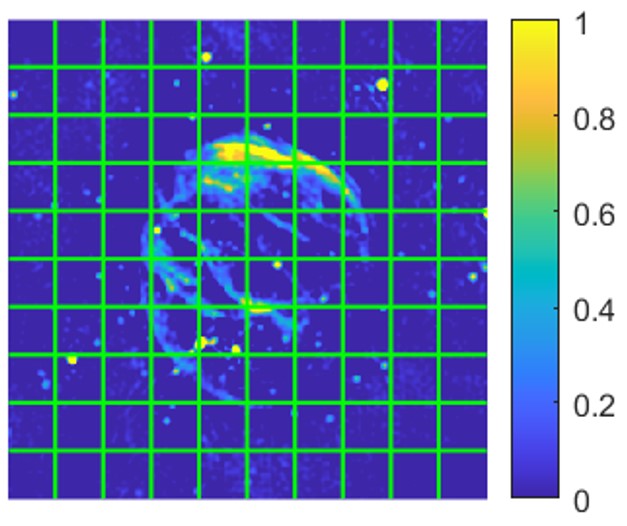}{3in}{(a)}
          \fig{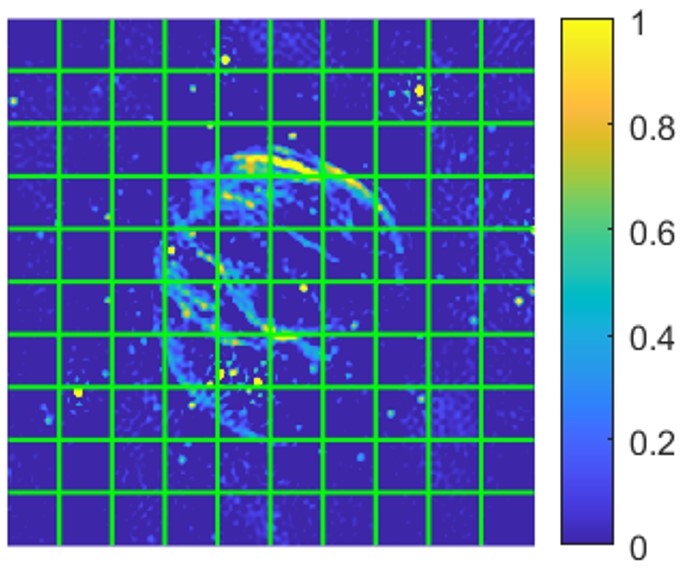}{3in}{(b)}
          }
\caption{Regular tiles of images generated by (a) MS-CLEAN (in WSClean) and (b) IUWT on SNR\_G55\_10s. As shown in the colour scales, the intensity of each pixel has been normalised by the maximum value of the two images. The tile size can be adjusted based on users' requirement. The tile index (1,1) starts from the top-left corner of each image.
\label{tiles}}
\end{figure*}

Better still, IQA methods can work jointly without extraction of source(s) to analyse where the structure of the extended source(s) differs in the two input images to be compared. Take the analysis of MS-CLEAN vs IUWT of SNR\_G55\_10s as an example. The first step of the joint analysis should be to split the images into regular tiles, as shown in Fig. \ref{tiles}, so that the differences can be localised to each tile, also allowing the following computations on each tile to be performed in parallel. Calculate SSIM and augLISI on every pair of tiles formed from the same tile position in the two input images, to obtain values for the SSIM matrix and augLISI matrix. Note that SSIM is used because it does not have a preference regarding high- or low-intensity differences and augLISI is used as these images contain extended sources.

When comparing the calculated IQA values for each image tile, it is important to assess two quantities. The first is whether the difference between the IQA values is sufficiently small, where a threshold $\delta$ is chosen based on requirements, i.e., if
\begin{equation}
|\mathrm{augLISI}(I_1,I_2) - \mathrm{SSIM}(I_1,I_2)| \leq \delta,
\end{equation}
where $I_1$ and $I_2$ are inputs to be compared by the IQA indexes. The second is defining whether augLISI is sufficiently large, where a threshold $\tau$ is again chosen based on requirements, i.e., if
\begin{equation}
\mathrm{augLISI}(I_1,I_2) \geq \tau .
\end{equation}

\begin{figure}[!t]
	\centering
\includegraphics[width=3in]{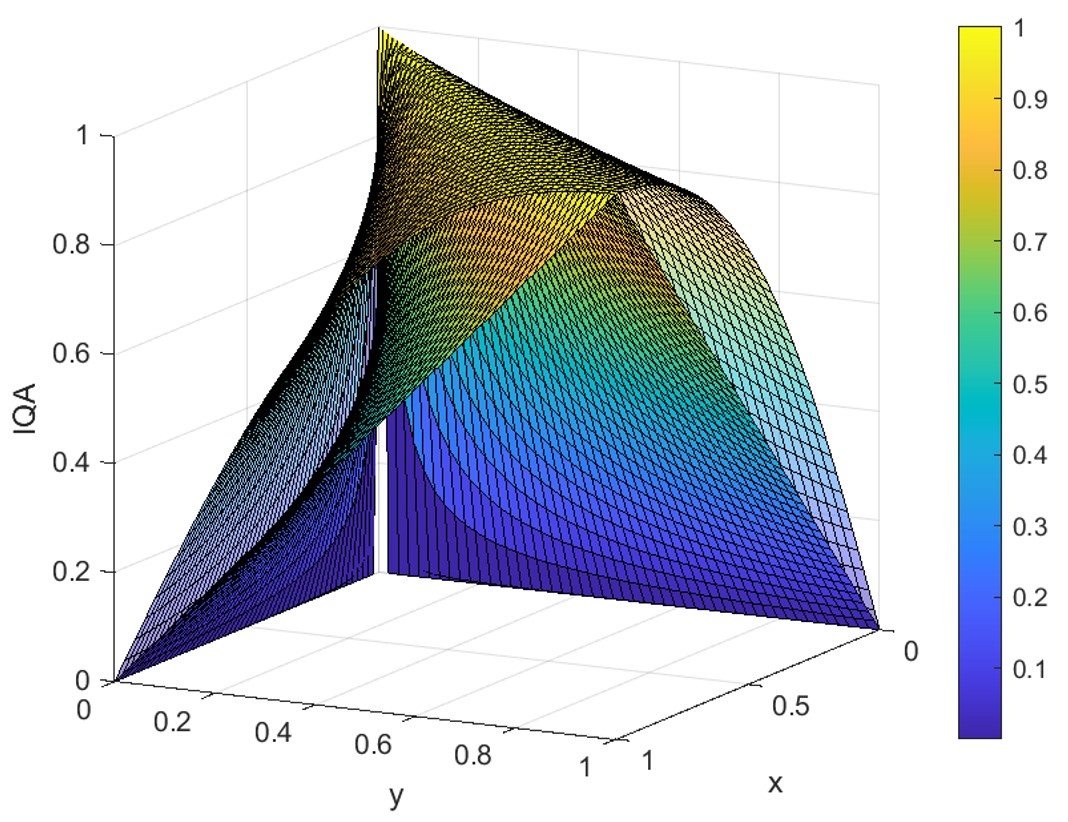}
\caption{Response of augLISI and SSIM to a pair of single-pixel images $(x,y)$, where intensity $x \in [0,1]$ and intensity $y \in [0,1]$. It shows augLISI and SSIM on a grid with a colour scale, where the surface with 50\% transparency represents the response of SSIM and the other surface represents the response of augLISI.
\label{aug3D}}
\end{figure}

To find meaningful values of $\delta$ and $\tau$, a simulation can be done based on the augLISI equation, i.e., Equation (\ref{augLISI}). To illustrate the response of augLISI and compare it to the response of SSIM, we generate two input images $x$ and $y$. Each image consists of a single pixel with an intensity value between 0 and 1. Fig. \ref{aug3D} shows the output of augLISI and SSIM for different $(x,y)$ ($x \in [0,1]$, $y \in [0,1]$).

\begin{figure*}
\gridline{\fig{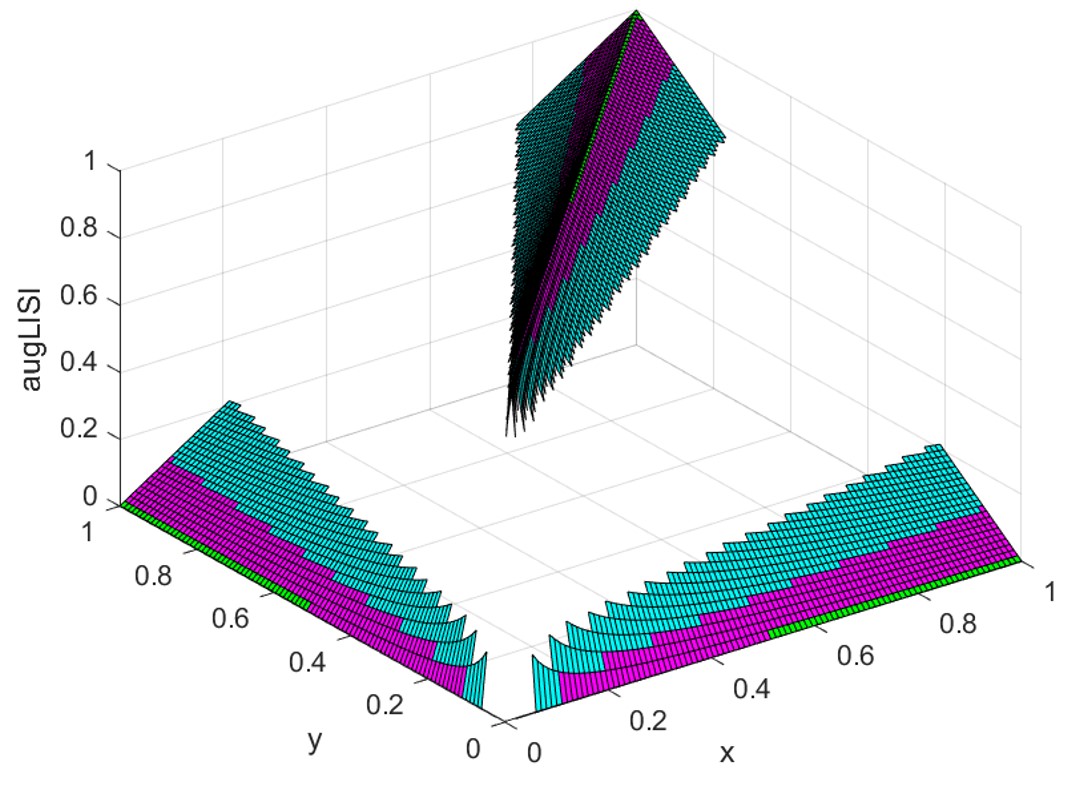}{3in}{(a)}
          \fig{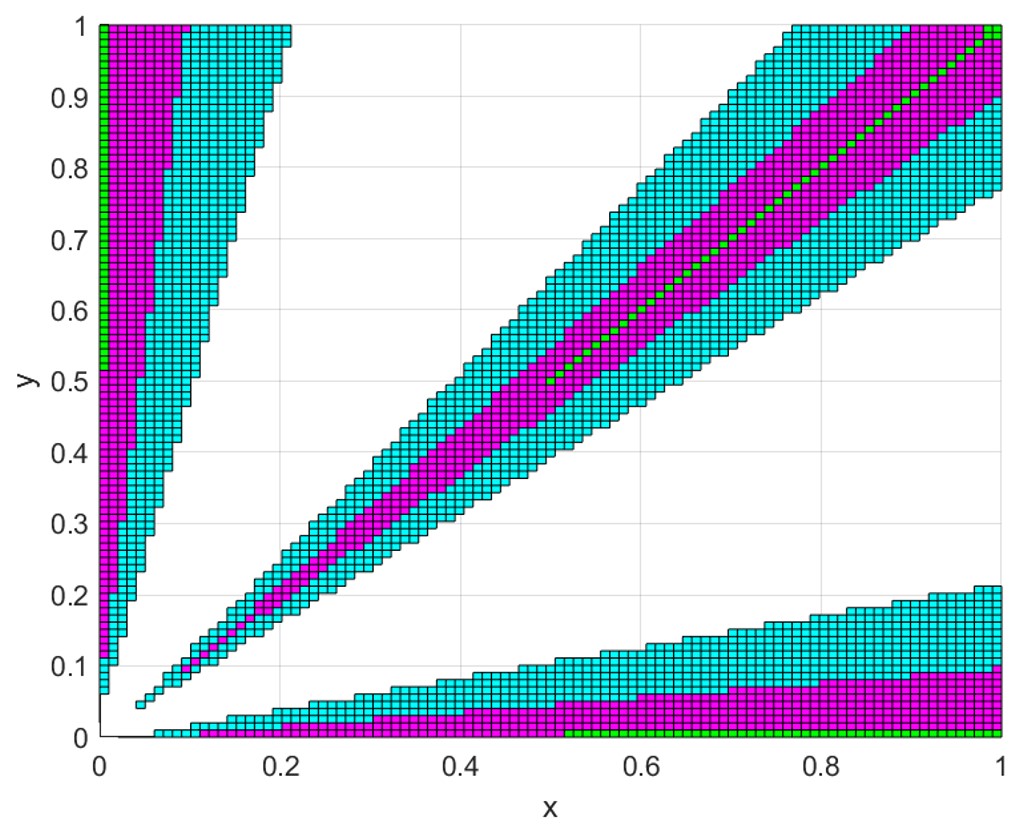}{3in}{(b)}
          }
\caption{Response of augLISI to pixel intensities for condition $|\mathrm{augLISI}(x,y) - \mathrm{SSIM}(x,y)| \leq \delta$. Different colours represent different values of $\delta$, where green points are for $\delta = 0.02$, magenta points are for $\delta = 0.1$, and cyan points are for $\delta = 0.2$. In this figure, (a) shows the 3D plot of the IQA results and (b) shows the 2D version of it by looking from the top of the 3D plot.
\label{closeenough}}
\end{figure*}

To further illustrate the difference between augLISI and SSIM for the same inputs, Fig. \ref{closeenough} shows regions where $|\mathrm{augLISI}(x,y) - \mathrm{SSIM}(x,y)| \leq \delta$. In other words, samples of inputs which have similar responses of augLISI and SSIM are shown in this figure, where the 'similar' is defined by $\delta$. There are two structures visible for each $\delta$. One is a cluster of IQA points with values approaching 1, where $x$ and $y$ are equal or close to each other. The other is the IQA points with values near 0 located on both $x$ and $y$-axes, where $x$ and $y$ are very different. In the areas without any points, the augLISI response is lower than the SSIM response, for the single-pixel inputs. The key difference between augLISI and SSIM is in the treatment of low-intensity values as demonstrated by the tapered shape of the two structures shown in Fig. \ref{closeenough}. The augLISI assigns low-intensity pixels, that are likely to be noise, small significance and thus it is less sensitive to their contribution when comparing images.

\begin{figure*}
\gridline{\fig{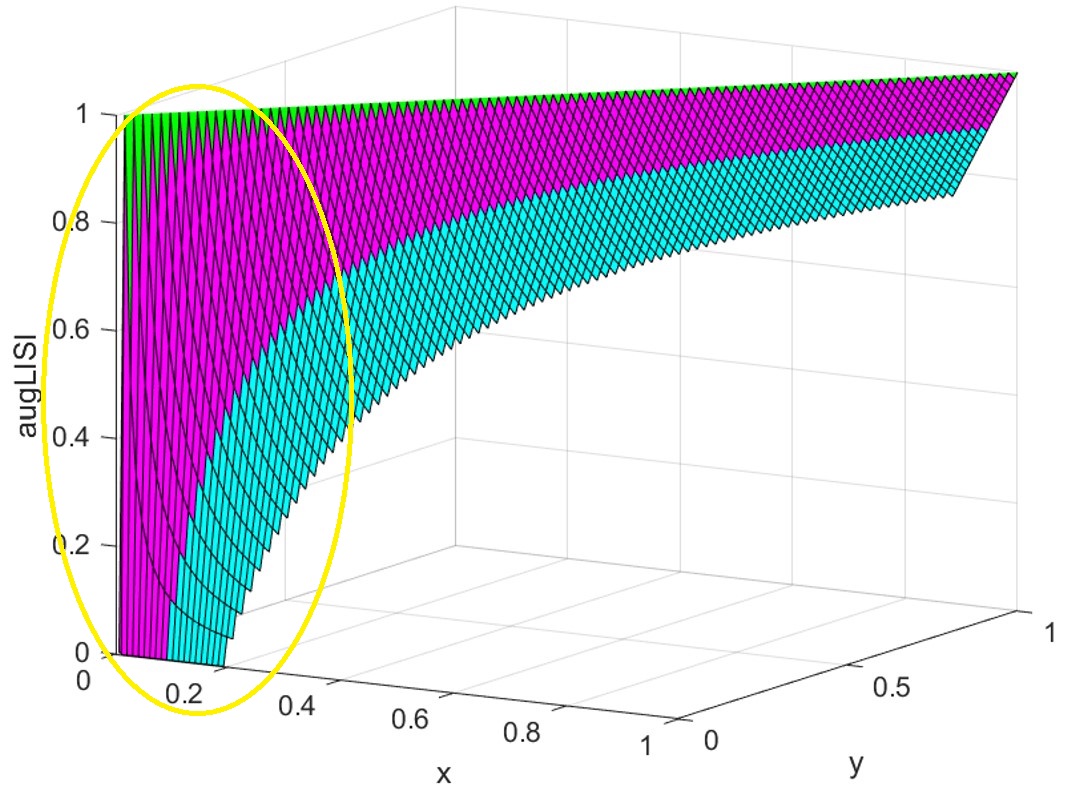}{3in}{(a)}
          \fig{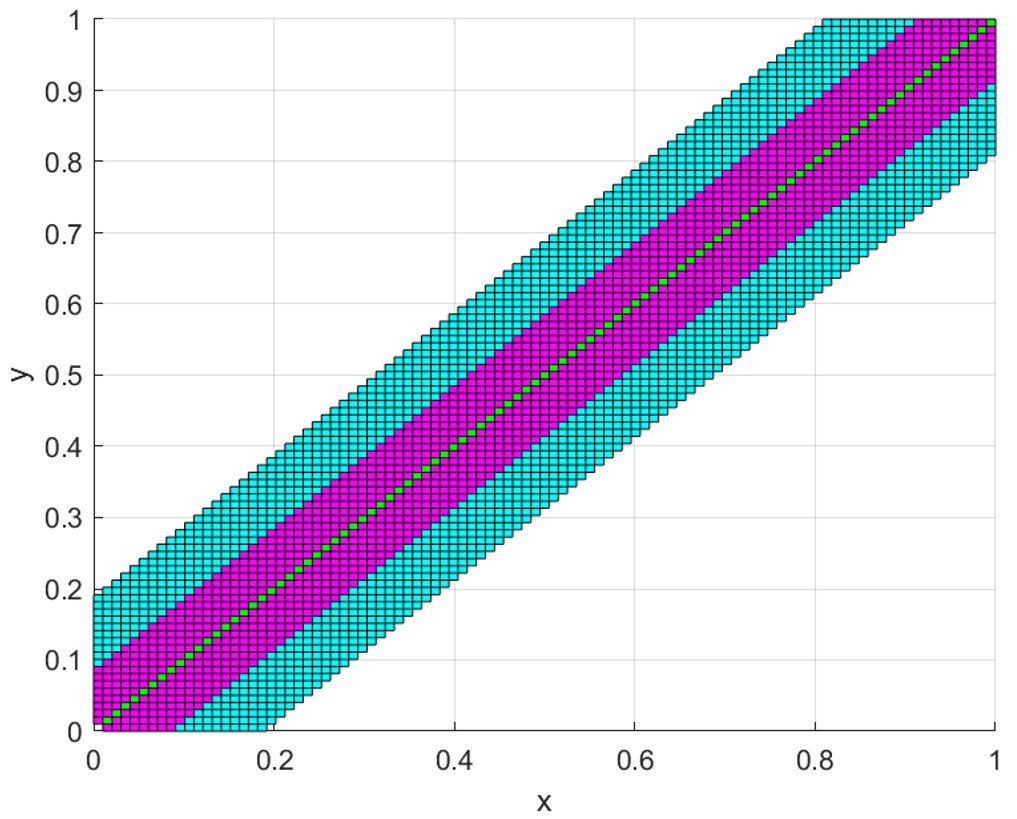}{3in}{(b)}
          }
\caption{Response of augLISI to pixel intensities $x$ and $y$ when $|x-y|$ is small. Sensitivity to low or near-zero intensities is suppressed in augLISI, so even similar values of $(x,y)$ do not definitely result in a response of 1. Different colours represent different differences between $x$ and $y$, where $|x-y| \leq 0.02$ for green points, $|x-y| \leq 0.1$ for magenta points, and $|x-y| \leq 0.2$ for cyan points. In this figure, (a) shows the 3D plot of the augLISI results and (b) shows the 2D version of it by looking from the top of the 3D plot. The yellow circle in (a) shows the regions of small intensities (noise).
\label{001}}
\end{figure*}

Smaller significance of low-intensity pixels also implies, as shown by the tip of the tapers in Fig. \ref{closeenough}, that low-intensity pixels must be very close to each other to be considered the same. In greater detail, this behaviour is illustrated in Fig. \ref{001}, where we see that the augLISI response can be smaller than 1 even for inputs with similar intensities when they are both small (yellow circle). The reason is that in this case $x$ and $y$ are very likely to be noise, while augLISI focuses on the similarity of sources and does not care about the similarity of noise.

We can determine the appropriate value of $\delta$ based on the relationship between $\delta$ (Fig. \ref{closeenough}) and $|x-y|$ (Fig. \ref{001}). The relationship between them is that the maximum difference between $x$ and $y$ for $\delta=0.02$ is 0.0200, for $\delta = 0.1$ is 0.1062, and for $\delta = 0.2$ is 0.2345. Therefore, $\delta$ needs to be set to satisfy the noise sensitivity required in users' application. The ranges of noise are different for different $\delta$, as shown Fig. \ref{001} (yellow circle). In our example for assessing the images generated by MS-CLEAN (in WSClean) and IUWT (Fig. \ref{tiles}), the noise distributions of these two figures are shown in Fig. \ref{noise} (bottom-left) and (top-left). The standard deviations $\sigma$ of the noise in the images generated by MS-CLEAN (in WSClean) and IUWT are 0.0201 and 0.0153, respectively. Therefore, based on the aforementioned relationship, we set $\delta = 0.02$ (a value with proper fractional precision closest to both standard deviations of noise) for assessing images of SNR\_G55\_10s.

\begin{figure*}
\gridline{\fig{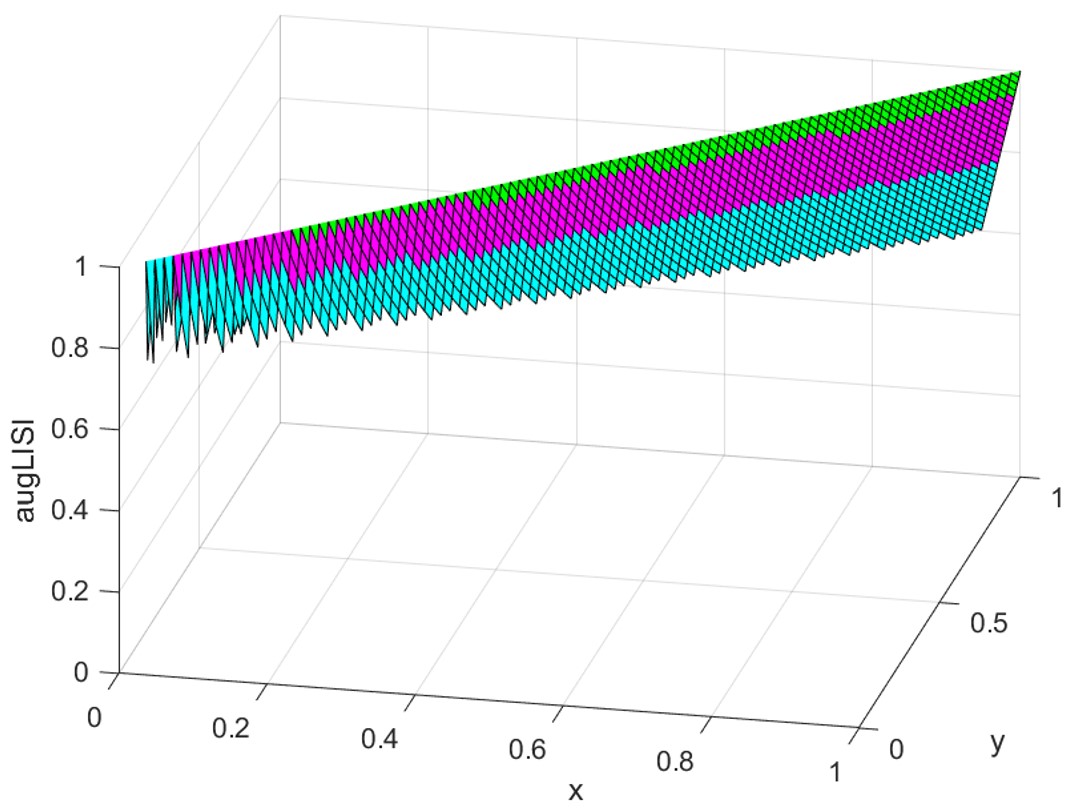}{3in}{(a)}
          \fig{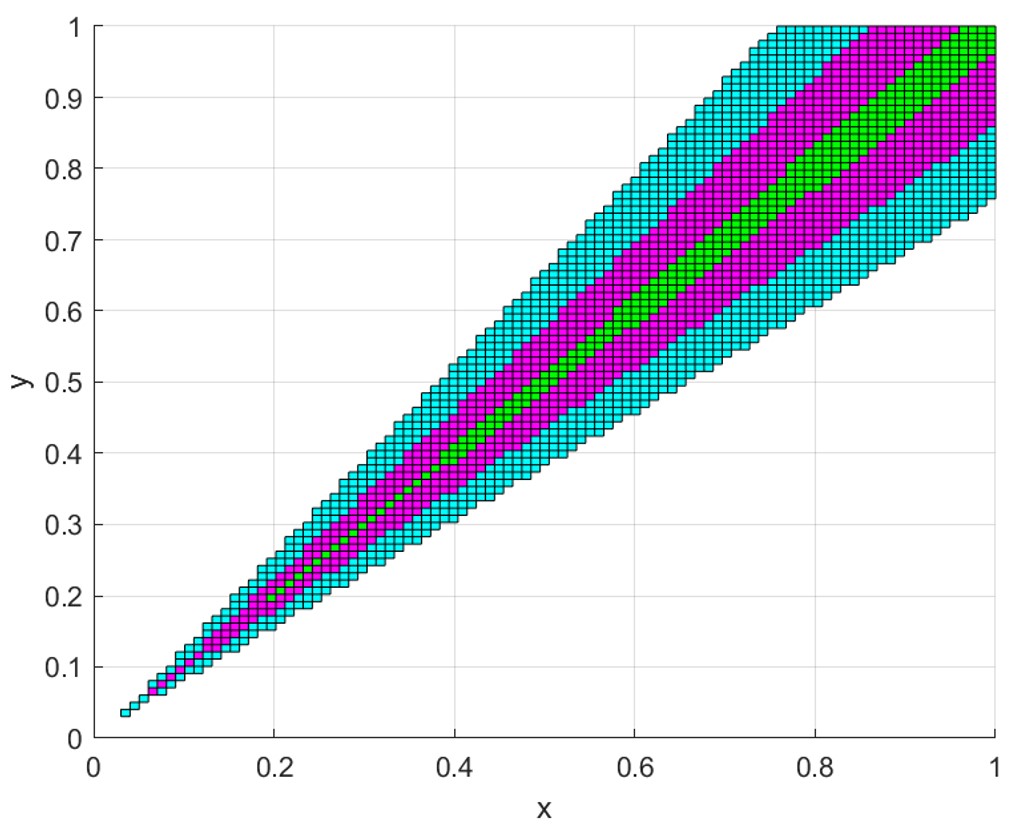}{3in}{(b)}
          }
\caption{Response of augLISI for $(x,y)$ with different $\tau$, where $\tau = 0.95$ for green points, $\tau = 0.85$ for magenta points, and $\tau = 0.75$ for cyan points. (a) shows the 3D plot of the augLISI results and (b) shows the 2D version of it by looking from the top of the 3D plot.
\label{085}}
\end{figure*}

To establish whether augLISI is sufficiently large, i.e., to determine $\tau$, users need to consider their required similarity between the images. Specifically, Fig. \ref{085} depicts the augLISI results for $(x,y)$ when setting different values of $\tau$. When $\tau$ is smaller, the tolerance to difference in larger intensities (i.e., large $x$ and $y$) is higher, reflected in the extended area of intensities shown Fig. \ref{085}, which means the sensitivity to difference in large intensity is lower. The tolerance/sensitivity to difference in small intensity does not change significantly. To select a proper $\tau$, the sensitivity can be approximated by the proportion of the coloured area to the area of the whole normalised region. In our example for assessing the images generated by MS-CLEAN (in WSClean) and IUWT (Fig. \ref{tiles}), we select $\tau = 0.85$ to keep the sensitivity of augLISI focused on the highest 15\% of intensities.

\begin{figure}[!t]
	\centering
\includegraphics[width=7in]{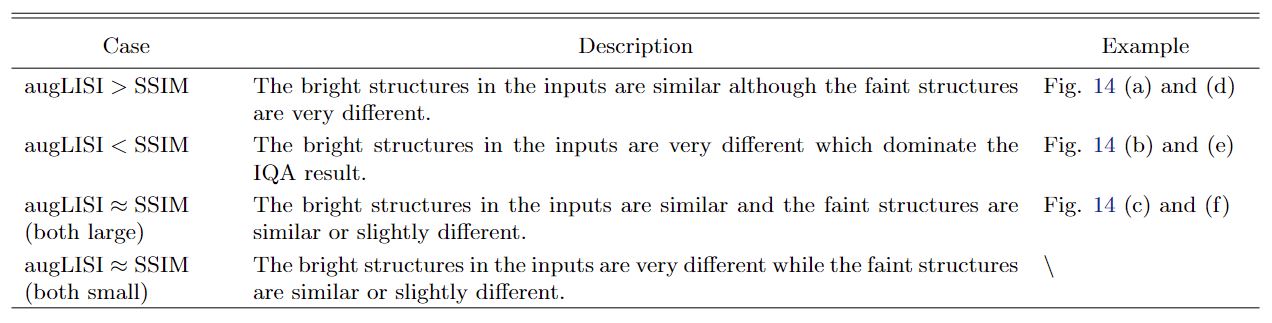}
\caption{Cases of augLISI and SSIM in joint analysis. This can be evidenced by considering the response to different input values into Equation (\ref{SSIM}) and (\ref{augLISI}).\label{jointaugssim}}
\end{figure}

Applying $\delta$ and $\tau$, generate four possible scenarios when comparing augLISI and SSIM which are shown in Table \ref{jointaugssim} (shown as a screenshot in the pre-print version). In the table, $\mathrm{augLISI} > \mathrm{SSIM}$ or $\mathrm{augLISI} < \mathrm{SSIM}$ indicates that the difference between augLISI and SSIM is more than $\delta$ ($\delta = 0.02$ in our example); $\mathrm{augLISI} \approx \mathrm{SSIM}$ indicates that the difference between augLISI and SSIM is smaller than $\delta$. The ``both large/small'' indicates that both augLISI and SSIM are larger/smaller than $\tau$ ($\tau = 0.85$ in our example). There is no tile in the example images, where $\mathrm{augLISI} \approx \mathrm{SSIM}$ (both small).
\begin{figure*}
\gridline{\fig{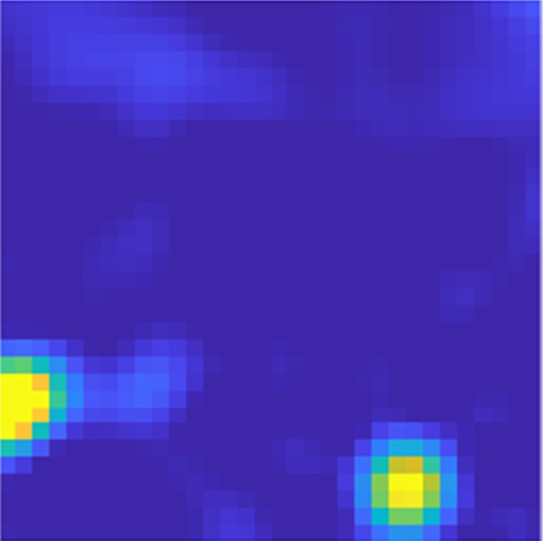}{1.6in}{(a)}
          \fig{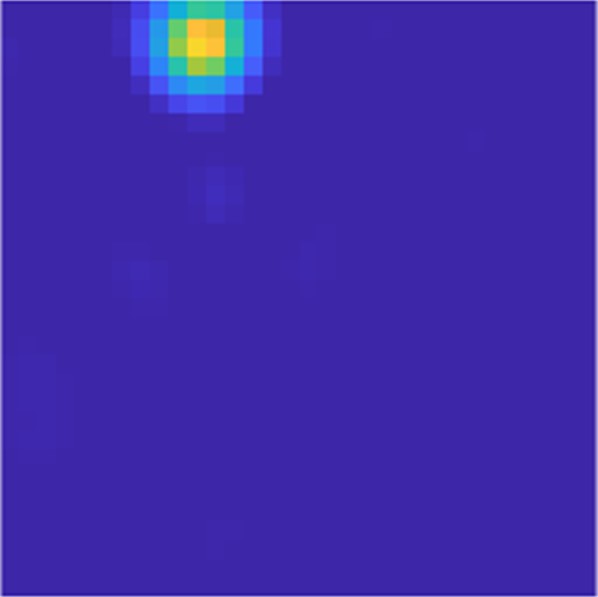}{1.6in}{(b)}
          \fig{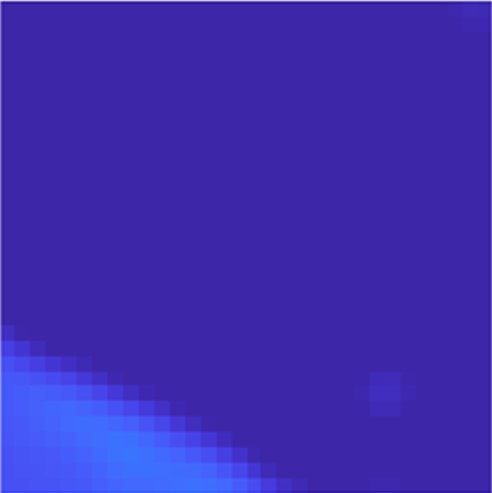}{1.6in}{(c)}
          }
\gridline{\fig{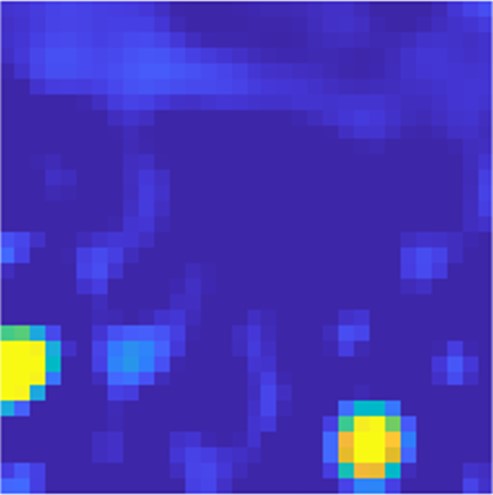}{1.6in}{(d)}
          \fig{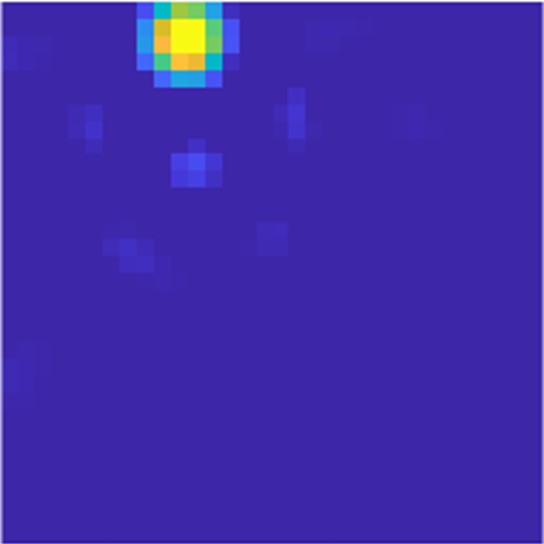}{1.6in}{(e)}
          \fig{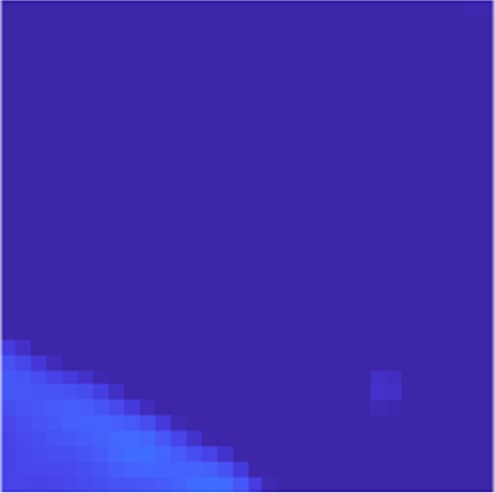}{1.6in}{(f)}
          }
\caption{With data of SNR\_G55\_10s, tile index (7,5), (8,2), and (3,7) in the images generated by MS-CLEAN are illustrated in (a), (b), and (c), respectively; those three tiles in the images generated by IUWT are illustrated in (d), (e), and (f), respectively. The SSIM and augLISI values for each pair of tiles are shown in Table \ref{data14}.
\label{joint}}
\end{figure*}
\begin{deluxetable*}{ccc}
\tablecaption{SSIM and augLISI for assessing tile (7,5), (8,2), and (3,7)\label{data14}}
\tablewidth{0pt}
\tablehead{
\colhead{Tile index} & \colhead{SSIM} & \colhead{augLISI}
}
\startdata
(7,5) & 0.8169 & 0.9312\\
(8,2) & 0.9729 & 0.9405\\
(3,7) & 0.9878 & 0.9862\\
\enddata
\end{deluxetable*}

\begin{figure*}
\gridline{\fig{ 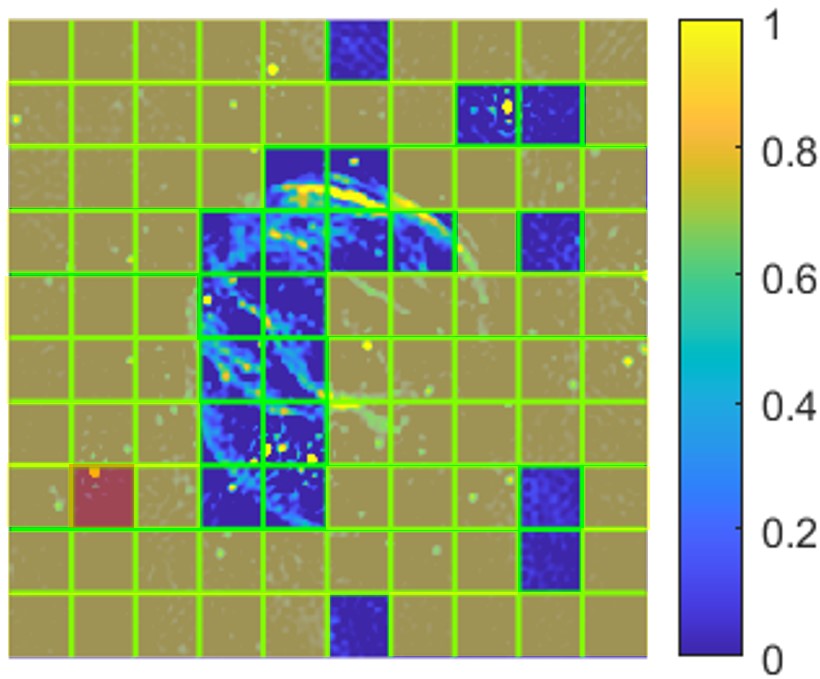}{3in}{(a)}
          \fig{ 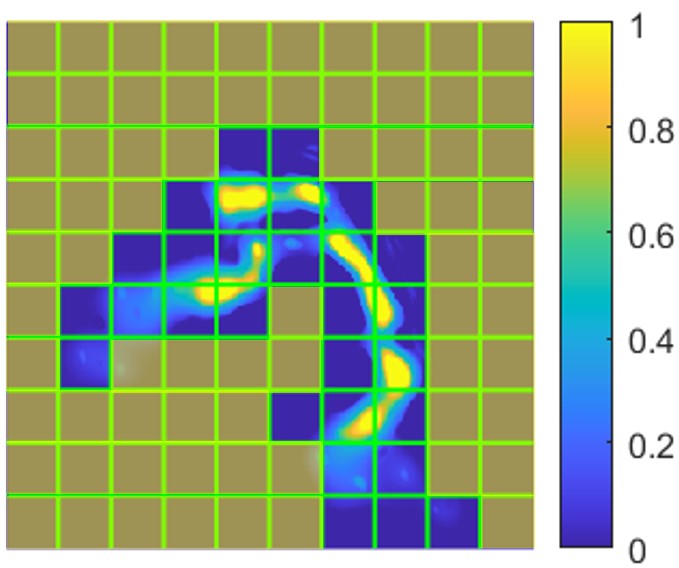}{3in}{(b)} }
\caption{Conclusions of comparing the images generated by MS-CLEAN (in WSClean) and IUWT for (a) SNR\_G55\_10s and (b) 3C75. In this figure, only the images generated by IUWT are illustrated to show the cases of different tiles. According to the results of joint analysis, in the yellow shadowed tiles, the bright structures in the images generated by MS-CLEAN and IUWT are similar and the faint structures are similar or slightly different; in the non-shadowed tiles, the bright structures in the images generated by MS-CLEAN and IUWT are similar although the faint structures are very different; and in the orange shadowed tile, the bright structures in the images generated by MS-CLEAN and IUWT are very different.
\label{gridded_summary}}
\end{figure*}

According to the IQAs of each tile in SNR\_G55\_10s, non-shadowed tiles in Fig. \ref{gridded_summary} (a) show that the bright structures in the inputs are similar although the faint structures are very different. Only tile (8,2) (the orange shadowed tile in Fig. \ref{gridded_summary} (a)) shows that the bright structures in the inputs are very different. Although there is no gold standard between the images generated by MS-CLEAN (in WSClean) and IUWT, the joint analysis suggests that further development/improvement of either algorithm should focus on the orange shadowed tile, which, in this example, includes a point-like source.

Similarly, to assess the images of 3C75 generated by MS-CLEAN (in WSClean) and IUWT, we select $\delta = 0.03$ and $\tau = 0.85$. As shown in Fig. \ref{gridded_summary} (b), it shows that the bright structures of the source are correctly restored but the faint structures of the source are very different. Further development/improvement of either algorithm (especially CLEAN) should focus on the fainter regions of this data.

Therefore, SSIM and augLISI can work jointly on RA images with extended sources, to characterise the differences in different parts of the source(s) structure without using source extraction.

\section{Conclusion}
RA images of the same SBD are known to be very similar, especially when they are observed by the HVS. We propose a new IQA method augLISI in this article, to assess the quality of extended source images. The IQA methods are applied to real datasets and simulated datasets in our experiments to illustrate how the IQA method can help radio astronomers with the study of RI imaging techniques.

AugLISI helps astronomers to compare images quantitatively, not only revealing the differences between different implementation approaches for the same imaging algorithm, but also assessing the differences between different algorithms. To further characterise the similarity of different sections in the images, SSIM and augLISI can be used to assess the image quality jointly.

In this article, we focus on assessing the quality of radio astronomical images. However, our proposed IQA methods could also be applied to other astronomical areas of study. In the future work, our IQA methods may also contribute to automatic comparison of images for quality assessment, for example, in Continuous Integration/Continuous Delivery (CI/CD) pipeline \citep{cicd}. Furthermore, IQA methods can also help with transient search \citep{transearch}, for example, supernova hunting \citep{suphun}, by comparing relevant images to look for transients without source extraction.

\begin{acknowledgments}
This work has received support from STFC Grant (ST/W001969/1). The National Radio Astronomy Observatory is a facility of the National Science Foundation operated under cooperative agreement by Associated Universities, Inc. The authors would like to thank Fred Dulwich and Ben Mort for valuable guidance on OSKAR. The authors would also like to thank Dan Walker, Hannah Stacey, Dirk Petry, and Katharina Immer from ALMA Science help center for helpful advice regarding CASA. Also, the authors would like to thank Miguel Cárcamo for beneficial discussion about GPUVMEM. The authors would like to thank Radostin Stoyanov for his kind help of installing GPUVMEM.
\end{acknowledgments}

\clearpage
\appendix

\section{ITW-SSIM}
\label{itwsection}
We define a weighted mean, by weighting each pixel with its importance and summing over all pixels, expressed as
\begin{equation}
\label{mean}
    {\overset{-}{X}}_{N} = {\sum\limits_{i = 1}^{N}{f\left( X_{i} \right)X_{i}}},
\end{equation}
where $X$ is the value of a specific pixel and $i$ is the index of the pixel (where the maximum value of $i$ is $N$), and $f(X_{i})$ is the weighting factor for the pixel at location $i$.

Also mathematically, the weighted variance and weighted cross covariance are expressed as 
\begin{equation}
\label{var}
    \sigma^{2} = \frac{1}{N - 1}{\sum\limits_{i = 1}^{N}\left( f\left( X_{i} \right)NX_{i} - {\overset{-}{X}}_{N} \right)^{2}}
\end{equation}
and
\begin{equation}
\label{covar}
    \sigma_{XY} = \frac{1}{N - 1}{\sum\limits_{i = 1}^{N}{\left( f\left( X_{i} \right)NX_{i} - {\overset{-}{X}}_{N} \right)\left( f\left( Y_{i} \right)NY_{i} - {\overset{-}{Y}}_{N} \right)}},
\end{equation}
respectively, where $X$ and $Y$ are the values of a specific pixel at the same position of the two input images. It is known that the degree of freedom is $N-1$ in the original form of covariance \citep{DoF}. Hence the degree of freedom should also be $N-1$ in the weighted form, the full derivation and proof are presented in Appendix \ref{freedom}.

Reflecting in IQA, $X$ and $Y$ represent intensities of pixels of the two input images to be compared.

Thereby, the ITW-SSIMs (whose range should be $\left\lbrack - 1,1 \right\rbrack$) can be expressed by \begin{subequations}
\label{ITW}
\begin{equation}
\label{ITWa}
    \mathrm{ITW-SSIM}\left( {x,y} \right) = \frac{\left( {2\mu_{x}\mu_{y} + C_{1}} \right)\left( {2\sigma_{xy} + C_{2}} \right)}{\left( {\mu_{x}^{2} + \mu_{y}^{2} + C_{1}} \right)\left( {\sigma_{x}^{2} + \sigma_{y}^{2} + C_{2}} \right)}
\end{equation}
\begin{equation}
\label{ITWb}
\left\{ \begin{matrix}
{\mu_{x} = {\sum\limits_{i = 1}^{N}{f\left( x_{i} \right)x_{i}}}~~~~~~~~~~~~~~~~~~~~~~~~~~~~~~~~~~~~~~~~} \\
{\sigma_{x} = \left( {\frac{1}{N-1}{\sum\limits_{i = 1}^{N}\left( {f\left( x_{i} \right)Nx_{i} - \mu_{x}} \right)^{2}}} \right)^{\frac{1}{2}}~~~~~~~~~~~~~~~} \\
{\sigma_{xy} = \frac{1}{N-1}{\sum\limits_{i = 1}^{N}{\left( {f\left( x_{i} \right)Nx_{i} - \mu_{x}} \right)\left( {f\left( y_{i} \right)Ny_{i} - \mu_{y}} \right)}}} \\
\end{matrix} \right.
\end{equation}
\end{subequations}
where the inputs $x$ and $y$ are normalised, $N$ is the total number of pixels in the image, $\mu_{y}$ (or $\sigma_{y}$) is similar to $\mu_{x}$ (or $\sigma_{y}$), and the sum of weighting factors over a particular image should be 1, i.e., ${\sum\limits_{i = 1}^{N}{f\left( x_{i} \right)}} = 1$ and ${\sum\limits_{i = 1}^{N}{f\left( y_{i} \right)}} = 1$. The weighting functions of ITW-SSIMs can be selected by users for different applications. To weight higher intensity with higher importance, three weighting functions are implemented in ITW-SSIMs. They are Gaussian weighting, tanh weighting, and sigmoid weighting.

\section{Proof of the degree of freedom in ITW-SSIMs}
\label{freedom}
\begin{figure}[ht]
    \centering
    \includegraphics[width=7in]{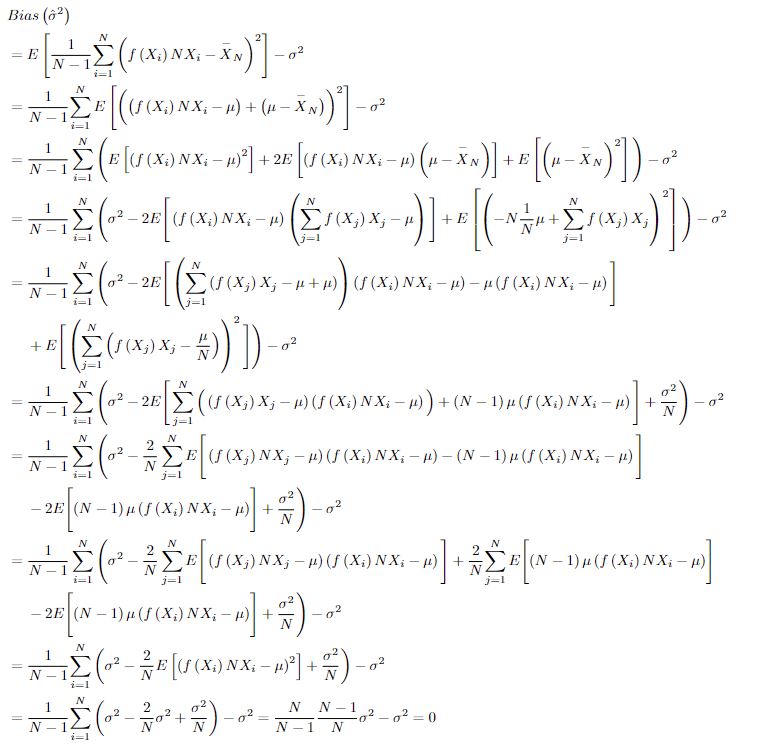}
\caption{Proof of the degree of freedom in ITW-SSIMs.}
\label{preprint}
\end{figure}

\noindent \textbf{\textit{Proof:}} To prove the degree of freedom is $N-1$ in weighted variance, bias of the estimator: (Shown as a screenshot in the pre-print version, please see Fig. \ref{preprint}.)

Because the bias equals to 0, the degree of freedom of weighted variance should be $N-1$.

The proof for weighted cross covariance is similar.

\noindent \textbf{\textit{Q.E.D.}}\\

\bibliographystyle{aasjournal}
\bibliography{Bibliography.bib}

\begin{thebibliography}{}
\expandafter\ifx\csname natexlab\endcsname\relax\def\natexlab#1{#1}\fi
\providecommand{\url}[1]{\href{#1}{#1}}
\providecommand{\dodoi}[1]{doi:~\href{http://doi.org/#1}{\nolinkurl{#1}}}
\providecommand{\doeprint}[1]{\href{http://ascl.net/#1}{\nolinkurl{http://ascl.net/#1}}}
\providecommand{\doarXiv}[1]{\href{https://arxiv.org/abs/#1}{\nolinkurl{https://arxiv.org/abs/#1}}}

\bibitem[{Ables(1974)}]{MEM9}
Ables, J. 1974, Astron. Astrophys. Suppl., 15, 383

\bibitem[{Asano \& Asoh(2011)}]{acoustics}
Asano, F., \& Asoh, H. 2011, in 2011 19th European Signal Processing Conference, 2009--2013

\bibitem[{Astro(2023)}]{cicd}
Astro. 2023, {Astro: CI/CD}, \url{https://academy.astronomer.io/astro-module-cicd}

\bibitem[{Bakurov {et~al.}(2022)Bakurov, Buzzelli, Schettini, Castelli, \& Vanneschi}]{BAKUROV2022116087}
Bakurov, I., Buzzelli, M., Schettini, R., Castelli, M., \& Vanneschi, L. 2022, Expert Systems with Applications, 189, 116087, \dodoi{https://doi.org/10.1016/j.eswa.2021.116087}

\bibitem[{Briggs(1995)}]{briggs}
Briggs, D.~S. 1995, PhD thesis, The New Mexico Institue of Mining and Technology

\bibitem[{Brunet(2012)}]{SSIM13}
Brunet, D. 2012, PhD thesis, University of Waterloo

\bibitem[{Burke {et~al.}(2019)Burke, Graham-Smith, \& Wilkinson}]{RAbook}
Burke, B., Graham-Smith, F., \& Wilkinson, P. 2019, {An Introduction to Radio Astronomy}, 4th edn. (Cambridge: Cambridge Univ. Press)

\bibitem[{Carrillo {et~al.}(2014)Carrillo, McEwen, \& Wiaux}]{RA6}
Carrillo, R., McEwen, J., \& Wiaux, Y. 2014, Mon. Not. R. Astron. Soc., 439, 3591

\bibitem[{CASA(2010)}]{extended_data}
CASA. 2010, G055.7+3.4, \url{https://casaguides.nrao.edu/index.php?title=VLA_CASA_Imaging-CASA5.7.0}

\bibitem[{CASA(2018)}]{smiledata}
---. 2018, {CASA Guides:Polarization Calibration based on CASA pipeline standard reduction: The radio galaxy 3C75-CASA6.1.2}, \url{https://casaguides.nrao.edu/index.php?title=CASA_Guides:Polarization_Calibration_based_on_CASA_pipeline_standard_reduction:_The_radio_galaxy_3C75-CASA6.1.2}

\bibitem[{Chen {et~al.}(2006)Chen, Yang, \& Xie}]{SSIM5}
Chen, G., Yang, C., \& Xie, S. 2006, in {2006 International Conference on Image Processing}, 2929--2932

\bibitem[{Cornwell(2008)}]{CLEAN2}
Cornwell, T. 2008, IEEE J. Sel. Topics Signal Process., 2, 793

\bibitem[{Cornwell \& Evans(1985)}]{MEM10}
Cornwell, T., \& Evans, K. 1985, A\&A, 143, 77

\bibitem[{Cárcamo {et~al.}(2018)Cárcamo, Román, Casassus, Moral, \& Rannou}]{GPUVMEM}
Cárcamo, M., Román, P., Casassus, S., Moral, V., \& Rannou, F. 2018, Astronomy and Computing, 22, 16

\bibitem[{Cárcamo {et~al.}(2022)Cárcamo, Scaife, Alexander, \& Leahy}]{csromer}
Cárcamo, M., Scaife, A. M.~M., Alexander, E.~L., \& Leahy, J.~P. 2022, Monthly Notices of the Royal Astronomical Society, 518, 1955

\bibitem[{Dabbech {et~al.}(2015)Dabbech, Ferrari, Mary, Slezak, Smirnov, \& Kenyon}]{RA3}
Dabbech, A., Ferrari, C., Mary, D., {et~al.} 2015, A\&A, 576

\bibitem[{de~Lera~Acedo {et~al.}(2016)de~Lera~Acedo, Faulkner, \& Bij~de Vaate}]{aa22}
de~Lera~Acedo, E., Faulkner, A., \& Bij~de Vaate, J. 2016, in 2016 United States National Committee of URSI National Radio Science Meeting (USNC-URSI NRSM), 1--2

\bibitem[{Dewdney {et~al.}(2013)Dewdney, Turner, Millenaar, McCool, Lazio, \& Cornwell}]{SKA1}
Dewdney, P., Turner, W., Millenaar, R., {et~al.} 2013, {SKA1 System Baseline Design}, Tech. Rep. SKA-TEL-SKO-DD-001, SKA Organisation, Macclesfield

\bibitem[{Ding {et~al.}(2021)Ding, Ma, Wang, \& Simoncelli}]{Ding2021}
Ding, K., Ma, K., Wang, S., \& Simoncelli, E.~P. 2021, International Journal of Computer Vision, 129, 1258, \dodoi{10.1007/s11263-020-01419-7}

\bibitem[{Dulwich(2020)}]{OSKAR1}
Dulwich, F. 2020, {OSKAR 2.7.6}, \url{https://zenodo.org/records/3758491}

\bibitem[{Fornasier \& Rauhut(2011)}]{RIP}
Fornasier, M., \& Rauhut, H. 2011, {Handbook of Mathematical Methods in Imaging} (Springer)

\bibitem[{Frieden(1972)}]{MEM7}
Frieden, B. 1972, J. Opt. Soc. Am., 62, 511

\bibitem[{Garsden {et~al.}(2015)Garsden, Girard, Starck, Corbel, Tasse, Woiselle, McKean, Van~Amesfoort, Anderson, Avruch, \& et~al.}]{CS15}
Garsden, H., Girard, J., Starck, J., {et~al.} 2015, A\&A, 575

\bibitem[{Gull \& Daniell(1978)}]{MEM2}
Gull, S., \& Daniell, G. 1978, Nature, 272, 686

\bibitem[{Hardy(2013)}]{CS19}
Hardy, S. 2013, A\&A, 557

\bibitem[{Hogbom(1974)}]{CLEAN1}
Hogbom, J. 1974, Astron. Astrophys. Suppl., 15, 417

\bibitem[{Humphreys \& Cornwell(2011)}]{w5}
Humphreys, B., \& Cornwell, T. 2011, SKA MEMO, 132

\bibitem[{IUWT(2011)}]{iuwt_code}
IUWT. 2011, Deconvolution methods PF \& IUWT a new reweighted version, \url{https://code.google.com/archive/p/csra/downloads}

\bibitem[{Kaghaz-Garan {et~al.}(2014)Kaghaz-Garan, Umbarkar, \& Doboli}]{acoustics2}
Kaghaz-Garan, S., Umbarkar, A., \& Doboli, A. 2014, in 2014 IEEE International Symposium on Robotic and Sensors Environments (ROSE) Proceedings, 49--54

\bibitem[{Lane(2008)}]{DoF}
Lane, D.~M. 2008, {Degrees of Freedom}, \url{https://davidmlane.com/hyperstat/A42408.html}

\bibitem[{Li {et~al.}(2011{\natexlab{a}})Li, Brown, Cornwell, \& de~Hoog}]{CS16}
Li, F., Brown, S., Cornwell, T., \& de~Hoog, F. 2011{\natexlab{a}}, A\&A, 531

\bibitem[{Li {et~al.}(2011{\natexlab{b}})Li, Cornwell, \& de~Hoog}]{CS14}
Li, F., Cornwell, T., \& de~Hoog, F. 2011{\natexlab{b}}, A\&A, 528

\bibitem[{Li {et~al.}(2024)Li, Adamek, \& Armour}]{zenodo}
Li, X., Adamek, K., \& Armour, W. 2024, Intensity Sensitive IQAs v1.0.0, \dodoi{10.5281/zenodo.10863656}

\bibitem[{Li \& Armour(2022)}]{Wider}
Li, X., \& Armour, W. 2022, in 2022 26th International Conference on Pattern Recognition (ICPR), 1975--1981

\bibitem[{Lu {et~al.}(2022)Lu, Hodosan, Wang, Daley-Yates, \& Cornwell}]{currentQA}
Lu, Y.-H.~C., Hodosan, G., Wang, F., Daley-Yates, S., \& Cornwell, T. 2022, in 2022 3rd URSI Atlantic and Asia Pacific Radio Science Meeting (AT-AP-RASC), 1--4

\bibitem[{McMullin {et~al.}(2007)McMullin, Waters, Schiebel, Young, \& Golap}]{CASA}
McMullin, J., Waters, B., Schiebel, D., Young, W., \& Golap, K. 2007, in {Astronomical Data Analysis Software and Systems XVI}, ed. R.~Shaw, F.~Hill, \& D.~Bell, Vol. 376 (ASP Conference Series), 127--130

\bibitem[{Moorthy \& Bovik(2009)}]{SSIM8}
Moorthy, A., \& Bovik, A. 2009, IEEE J. Sel. Top. Signal Process., 3, 193

\bibitem[{Narayan \& Nityananda(1984{\natexlab{a}})}]{MEM}
Narayan, R., \& Nityananda, R. 1984{\natexlab{a}}, in {Proc. IAU/URSI Symp}, ed. J.~Roberts (Cambridge Univ. Press), 267

\bibitem[{Narayan \& Nityananda(1984{\natexlab{b}})}]{MEM3}
Narayan, R., \& Nityananda, R. 1984{\natexlab{b}}, in {Indirect Imaging}, ed. J.~Roberts (Cambridge: Cambridge University Press), 281--290

\bibitem[{NRAO(latest)}]{casaweb}
NRAO. latest, {Common Astronomy Software Applications (CASA)}, \url{https://casa.nrao.edu/}

\bibitem[{Offringa {et~al.}(2014)Offringa, McKinley, Hurley-Walker, {et~al.}}]{WS1}
Offringa, A.~R., McKinley, B., Hurley-Walker, {et~al.} 2014, MNRAS, 444, 606

\bibitem[{Offringa \& Smirnov(2017)}]{WS2}
Offringa, A.~R., \& Smirnov, O. 2017, MNRAS, 471, 301

\bibitem[{Oppenheim \& Lim(1981)}]{HVS}
Oppenheim, A., \& Lim, J. 1981, Proceedings of the IEEE, 69, 529

\bibitem[{OxfordSKA(V2.9.4)}]{OSKAR}
OxfordSKA. V2.9.4, {OxfordSKA / OSKAR}, \url{https://github.com/OxfordSKA/OSKAR}

\bibitem[{Ponsonby(1973)}]{MEM8}
Ponsonby, J. 1973, Mon. Not. R. Astron. Soc., 163, 369

\bibitem[{Pratley {et~al.}(2018)Pratley, McEwen, d'Avezac, Carrillo, Onose, \& Wiaux}]{RA4}
Pratley, L., McEwen, J., d'Avezac, M., {et~al.} 2018, Mon. Not. R. Astron. Soc., 473, 1038

\bibitem[{Raveendran {et~al.}(2020)Raveendran, Singh, \& Kumar}]{SSIM2}
Raveendran, R., Singh, A., \& Kumar, R. 2020, CoRR, abs/2007.05853

\bibitem[{Sampat {et~al.}(2009)Sampat, Wang, Gupta, Bovik, \& Markey}]{SSIM3}
Sampat, M., Wang, Z., Gupta, S., Bovik, A., \& Markey, M. 2009, IEEE Trans. Image Process., 18, 2385

\bibitem[{SKAO(2021)}]{aa2}
SKAO. 2021, {Green light given for construction of world’s largest radio telescope arrays}, \url{https://www.skao.int/en/news/174/green-light-given-construction-worlds-largest-radio-telescope-arrays}

\bibitem[{Sun {et~al.}(2015)Sun, Landecker, Gaensler, Carretti, Reich, Leahy, N.~McClure-Griffiths, Wolleben, Haverkorn, Douglas, \& Gray}]{CS17}
Sun, X., Landecker, T., Gaensler, B., {et~al.} 2015, Astrophys. J., 811

\bibitem[{Teeninga {et~al.}(2016)Teeninga, Moschini, Trager, \& Wilkinson}]{faiextra}
Teeninga, P., Moschini, U., Trager, S.~C., \& Wilkinson, M.~H. 2016, Mathematical Morphology - Theory and Applications, 1

\bibitem[{The\_CASA\_Team {et~al.}(2022)The\_CASA\_Team, Bean, Bhatnagar, Castro, Meyer, Emonts, Garcia, Garwood, Golap, Villalba, \& et~al}]{CASAnew}
The\_CASA\_Team, Bean, B., Bhatnagar, S., {et~al.} 2022, Publications of the Astronomical Society of the Pacific, 134

\bibitem[{Thompson(1999)}]{RA7}
Thompson, A. 1999, in {Synthesis Imaging in Radio Astronomy II}, ed. G.~Taylor, C.~Carilli, \& R.~Perley, Vol. 180 (Astronomical Society of the Pacific Conference Series), 11--36

\bibitem[{Thompson {et~al.}(2017)Thompson, Moran, \& Swenson}]{RAbook2}
Thompson, A., Moran, J., \& Swenson, G. 2017, {Interferometry and Synthesis in Radio Astronomy}, 3rd edn. (Oxford: Wiley)

\bibitem[{Van~der Tol {et~al.}(2018)Van~der Tol, Veenboer, \& Offringa}]{WS3}
Van~der Tol, S., Veenboer, B., \& Offringa, A.~R. 2018, A\&A, 616, A27

\bibitem[{Vu(2015)}]{suphun}
Vu, L. 2015, Supernova Hunting with Supercomputers, \url{https://cs.lbl.gov/news-media/news/2015/supernova-hunting-with-supercomputers/}

\bibitem[{Wang {et~al.}(2005)Wang, Bovik, \& Sheikh}]{SSIM12}
Wang, Z., Bovik, A., \& Sheikh, H. 2005, in {Digital Video Image Quality and Perceptual Coding, Marcel Dekker Series in Signal Processing and Communications}, ed. H.~Wu \& K.~Rao

\bibitem[{Wang {et~al.}(2004)Wang, Bovik, Sheikh, \& Simoncelli}]{SSIM1}
Wang, Z., Bovik, A., Sheikh, H., \& Simoncelli, E. 2004, IEEE Trans. Image Process., 13

\bibitem[{Wang \& Li(2011)}]{SSIM10}
Wang, Z., \& Li, Q. 2011, IEEE Trans. Image Process., 20, 1185

\bibitem[{Wang {et~al.}(2003)Wang, Simoncelli, \& Bovik}]{SSIM4}
Wang, Z., Simoncelli, E., \& Bovik, A. 2003, in {The Thrity-Seventh Asilomar Conference on Signals, Systems Computers, 2003}, Vol.~2, 1398--1402

\bibitem[{Wenger {et~al.}(2010)Wenger, Magnor, Pihlstrom, Bhatnagar, \& Rau}]{CS18}
Wenger, S., Magnor, M., Pihlstrom, Y., Bhatnagar, S., \& Rau, U. 2010, Publ. Astron. Soc. Pac., 122, 1367

\bibitem[{Wernecke \& d'Addario(1977)}]{MEM6}
Wernecke, S., \& d'Addario, L. 1977, IEEE Trans. Comput., C-26, 351

\bibitem[{Wiaux {et~al.}(2009)Wiaux, Jacques, Puy, Scaife, \& Vandergheynst}]{iuwt}
Wiaux, Y., Jacques, L., Puy, G., Scaife, A. M.~M., \& Vandergheynst, P. 2009, Monthly Notices of the Royal Astronomical Society, 395, 1733

\bibitem[{Wright {et~al.}(2017)Wright, Lintott, Smartt, Smith, Fortson, Trouille, Allen, Beck, Bouslog, Boyer, Chambers, Flewelling, Granger, Magnier, McMaster, Miller, O'Donnell, Simmons, Spiers, Tonry, Veldthuis, Wainscoat, Waters, Willman, Wolfenbarger, \& Young}]{transearch}
Wright, D.~E., Lintott, C.~J., Smartt, S.~J., {et~al.} 2017, Monthly Notices of the Royal Astronomical Society, 472, 1315

\bibitem[{Ye {et~al.}(2021)Ye, Gull, Tan, \& Nikolic}]{w7}
Ye, H., Gull, S., Tan, S., \& Nikolic, B. 2021, High dynamic range wide field imaging method in radio interferometry

\bibitem[{Zhu {et~al.}(2019)Zhu, Liu, Wang, Zhang, Tian, Yu, Liang, Wu, Hu, \& Duan}]{contexextra}
Zhu, M., Liu, W., Wang, B.~Y., {et~al.} 2019, Advances in Astronomy, 2019

\bibitem[{Zhu {et~al.}(2020)Zhu, Huang, Hou, Du, \& Song}]{BigData11}
Zhu, Y., Huang, T., Hou, J., Du, S., \& Song, S. 2020, in {Big Data in Astronomy}, ed. L.~Kong, T.~Huang, Y.~Zhu, \& S.~Yu (Elsevier), 271--303

\end{thebibliography}

\end{document}